\def\lesssim{\mathrel{\hbox{\rlap{\hbox{\lower4pt\hbox{$\sim$}}}\hbox{$<$}}}}
\def\gtrsim{\mathrel{\hbox{\rlap{\hbox{\lower4pt\hbox{$\sim$}}}\hbox{$>$}}}}

\documentclass[]{spie}  %>>> use for US letter paper
\usepackage[dvips]{graphicx}
\title{Ultra high-energy neutrino at GZK energy: Z-WW showering in dark halo and tau airshowers
                     emerging from the Earth}
\author{Fargion D. \supit{a}\supit{b}
\skiplinehalf
\supit{a}1,Rome University La Sapienza, Ple.A.Moro,2, Rome, Italy \\
\supit{b}2,INFN Rome1, Rome, Italy }

\authorinfo{Further author information: (Send correspondence to Daniele Fargion)\\
E-mail: daniele.fargion@roma1.infn.it, Telephone: +39 06 4991 4287\\
Address: Physics Department,  Rome University 1,Ple.A.Moro
2,00185, Rome, Italy }
\begin{document}
  \maketitle

%%%%%%%%%%%%%%%%%%%%%%%%%%%%%%%%%%%%%%%%%%%%%%%%%%%%%%%%%%%%%
\begin{abstract}
Relic neutrino $\nu_{r}$ light masses clustering in Galactic and
Local Hot Dark Halos act as a beam dump calorimeter. Ultra High
Energy $\nu$, above ZeV, born by AGNs,GRBs  at cosmic edges,
overcoming the  Greisen, Zatsepin, Kuzmin (GZK) cut-off, may hit
near Z resonance and WW-ZZ channels energies: their showering
into nucleons  and $\gamma$ Ultra High Cosmic Ray (UHECR) fit
observed  data. Any tiny  neutrino mass splitting may reflect
into  ({\em a twin }) bump at highest GZK energy cut-off. The
lighter the neutrino masses the higher the Z-Showering cut-off.
The Z or WW,ZZ showering
%into UHE $\gamma$, and UHECR $p$ $\bar{p}$ and $n$
%$\bar{n}$,
 might explain a peculiar clustering in observed UHECR
spectra at $10^{19}$, $2\cdot 10^{19}$, $4 \cdot 10^{19}$ eV
found recently by AGASA.  Coincidence of clustered UHECR with
highest $\gamma$ BLac sources, originated either by neutral and
charged particles (Q=0,+1,-1) is well tuned to Z-Showering
Scenario. Additional prompt TeVs signals
%(by consequent synchrotron radiation of UHE electron pairs secondaries)
 occur offering a natural solution of growing
Infrared-TeV cut-off paradoxes related to distant TeV BLac
sources, while electromagnetic cascades tail may explain
correlation found with GeV-EGRET Sources. Such UHE $\nu$
Astrophysics might trace near GZK  energy
 into Horizontal Tau Air-Showers originated
by the UHE $\nu_{\tau}$ Earth-Skimming  in wide Crown Earth Crust
around the observer. These Upward and Horizontal ${\tau}$
Air-Showers \textbf{UPTAUS}, \textbf{HORTAUS}, test huge crown
target volumes either from high mountains as well as observing
from planes, balloons or satellites. The \textbf{HORTAUS} from
mountains measure crown masses at UHE $\nu$ EeVs energies
comparable to few $km^3$, while from satellites at orbit
altitudes, at GZK energies
 $E_{\nu}\geq 10^{19}$, their corresponding Horizontal Crown
 Masses  may even exceed $150$ $km^3$. The expected event rate may produce
 at least a dozen of event a year within Z-WW Showering model from a satellite altitude.

\end{abstract}

%>>>> Include a list of keywords after the abstract


%%%%%%%%%%%%%%%%%%%%%%%%%%%%%%%%%%%%%%%%%%%%%%%%%%%%%%%%%%%%%

%%%%%%%%%%%%%%%%%%%%%%%%%%%%%%%%%%%%%%%%%%%%%%%%%%%%%%%%%%%%%
\section{Introduction}

 Light Neutrinos ($\sim 0.1-3 eV$) clustering in Galactic, Local Hot Dark
  Halo, being an efficient calorimeter for ZeV $\nu$ offer the possibility to
overcome the Cosmic Black Body opacity ($\gtrsim 4 \cdot
10^{19}\,eV$) (GZK)  at highest energy cosmic ray astrophysics.
These rare events, being  nearly isotropic, are probably of cosmic
origin. They are very possibly originated by blazars Jets AGN
sources pointing their huge linear accelerators to us, as BLac; in
standard scenario if the UHECR are originally of hadronic nature
they must be  absorbed by the dragging friction of cosmic 2.75 K
BBR or by  the inter-galactic radio backgrounds (the GZK cut-off).
Indeed as it has been noted (K.Greisen,\cite{Greisen 1966}
Zat'sepin, Kuz'min \cite{Zat'sepin et al.1966} 1966), proton and
nucleons mean free path at E $> 5 \cdot 10^{19} \,EeV$ is less
than 30 $Mpc$ and asymptotically nearly ten $Mpc$; also gamma
rays at those energies have even shorter interaction length ($10
\,Mpc$) due to severe opacity by electron pair production via
microwave and radio background interactions (J.W.Elbert,
P.Sommers \cite{Elbert et all.1995}, 1995)(R.J.Protheroe and
Biermann \cite{Protheroe and Biermann 1997}, 1997. Nevertheless
these powerful sources (AGN, Quasars, GRBs) suspected to be the
unique source able to eject such UHECRs, are rare  at nearby
distances ($\lesssim 10 \div 20 \, Mpc$, as for nearby $M87$ in
Virgo cluster); moreover there are not nearby $AGN$ in the
observed UHECR arrival direction cone. Strong and coherent
galactic \cite{Protheroe and Biermann 1997} or extra-galactic
magnetic fields \cite{Farrar et al.2000}, able to bend such UHECR
(proton, nuclei) directions, are not really at hand. The needed
coherent lengths and strength are not easily compatible with
known cosmic magnetic fields \cite{Elbert et all.1995}. Finally
in this scenario the $ZeV$ neutrons born, by photo-pion proton
conversions on BBR, may escape the magnetic fields bending and
should keep memory of the arrival direction, leading to
(unobserved)  clustering toward the primary source (Fargion et
all 2001a \cite{Fargion et al.2001a}, \cite{Fargion et
al.2001b}2001b). Secondaries EeV photons (by neutral pion decays)
should also abundantly point and cluster toward the same nearby
$AGN$ sources (P.Bhattacharjee et all \cite{Bhattacharjee et
al.2000}2000),(J.W.Elbert, P.Sommers \cite{Elbert et all.1995},
1995), in disagreement with $AGASA$ data (for any direct UHECR.
Therefore Galactic origin for UHECR might be imagined as a
simplest solution (by Micro-Quasars sources), but it contradicts
the absence of any evident quadruple (galactic plane) or dipole
(galactic halo) UHECR an-isotropy. A often revived solution of
the present GZK puzzle, the Topological defects ($TD$), assumes
as a source, relic heavy particles (GUT masses) of early
Universe; they are imagined diffused as a Cold Dark Matter
component, in galactic halo, but therefore they are unable to
explain the growing evidences of clustering in $AGASA$~ $UHECR$
arrival data and their self-correlation  with far Compact Blazars
(BLac) at cosmic distance  (Tinyakov P.G.et Tkachev
\cite{Tinyakov-Tkachev2001}2001; D.S.Gorbunov, P.G.Tinyakov,
I.I.Tkachev, S.V.Troitsky \cite{Gorbunov et al.2002} 2002). In
this frame work it is important to remind the Fly's Eye event
($300$ EeV) whose association with Seyfert Galaxy MCG 8-11-11 (or
Quasar 3C147), inspired  earliest articles (\cite{Fargion Salis
1997}D.Fargion, B.Mele, A.Salis \cite{Fargion Mele Salis
1999}1997-99) to solve GZK by Z-Shower. Therefore the solution of
UHECR puzzle based on primary Extreme High Energy (EHE) neutrino
beams (from AGN) at ZeV $E_{\nu} > 10^{21}$ eV and their
undisturbed propagation from cosmic distances up to nearby
calorimeter, made by relic light $\nu$ in dark galactic or local
dark halo (\cite{Fargion Salis 1997}D.Fargion, B.Mele, A.Salis
\cite{Fargion Mele Salis 1999}1997-99,Weiler \cite{Weiler1999}
1999,  S.Yoshida, G. Sigl, S. Lee \cite{Yoshida et al.1998}1998)
remains the most favorite convincing solution for the GZK puzzle.
New complex scenarios for each neutrino mass spectra are then
opening and important signatures of Z,WW showering must manifest
in observed an-isotropy, composition, spectra shape and
space-time clustering of present and future UHECR data.
\subsection{ UHE $\nu$  Astronomy Energy Windows}
    Rarest TeVs gamma signals are at present the most extreme and
    rarest trace of  High Energy Astrophysics. The TeVs signals have shown new
power-full Jets blazing to us from Galactic or extragalactic
edges. At PeVs energies astrophysical Gamma cosmic rays should
also be present, but, excluded a very rare and elusive Cyg$X3$
event, they are  not longer being observed. While the
corresponding PeVs charged cosmic rays are abundantly hitting the
atmosphere, these missing PeVs gamma sources are very probably
mostly absorbed by their photon interactions (photo-pion
productions, electron pairs creation) at the source environment
and/or along the photon propagation into the cosmic Black Body
Radiation (BBR) or into other diffused (radio,infrared,optical)
background radiation. Unfortunately PeVs charged cosmic rays,
bend and bounded in a  random walk by Galactic magnetic fields,
loose their original directionality and their astronomical
relevance; their resident time in the galaxy is much longer ($\geq
10^{3}$-$10^{5}$) than neutral ones, as gamma rays, making the
charged cosmic rays more probable to be observed by nearly a
comparable ratio. On the contrary astrophysical UHE neutrino
signals at $10^{13}$eV-$10^{19}$eV (or higher GZK energies) are
unaffected by any radiation cosmic opacity and may open a very
new exciting window to High Energy Astrophysics. Lower energy
astrophysical UHE $\nu$ at $10^{9}$eV-$10^{12}$eV should also be
present, but their signals are (probably) drowned  by the dominant
diffused atmospheric $\nu$ secondaries noises produced by the
same charged (and smeared) UHE cosmic rays (while hitting
terrestrial atmosphere), the so called atmospheric neutrinos. In
a very far corner, at lowest (MeVs) energy  windows, the abundant
and steady solar neutrino flux  and the prompt (but rarer)
neutrino burst from a nearby Super-Novae (SN 1987A), have been in
last twenty years, already successfully explored. The UHE
$10^{13}$eV-$10^{16}$eV $\nu$ 's  astronomy, being weakly
interacting and rarer, may  be captured mainly inside huge
volumes, bigger than Super-Kamiokande ones; at present  most
popular detectors consider underground ones (Cubic Kilometer Size
like AMANDA-NESTOR) or (at higher energy $10^{19}$eV-$10^{21}$eV)
the widest Terrestrial atmospheric sheet volumes (Auger-Array
Telescope or EUSO atmospheric Detectors). Underground $km^3$
detection is based mainly on $\nu_{\mu}$ (above hundred TeVs
energies, after their interaction with matter) leading to $\mu$
kilometer size lepton tracks \cite{Gandhi et al. 1998} . Rarest
atmospheric horizontal shower are also expected by $\nu$
interactions in air (and, as we shall discuss, in the Earth
Crust). While $km^3$ detectors are optimal for PeVs neutrino
muons, the Atmospheric Detectors (AUGER-EUSO like) exhibit a
minimal threshold at highest ($\geq 10^{18} eV$) energies.
%%%%%%%%%%%%%%%%%%%%%%%%%%%%%%%%%%%%%%%%%%%%%%%%%%%%%%%%%%%%%%%%%%%%%%
\section{Relic $\nu_r$ neutrino masses and Z-WW Channels and  Showering}
If relic neutrinos have a mass much  larger than their average
thermal energy (1.9 $K^0$ $\sim 5\cdot 10^{-4}$ eV) they may
cluster in galactic or Local Group halos; at eVs masses the
clustering seem very plausible and it may play a role in dark hot
cosmology\cite{Fargion 1983}. The relevant role in astrophysics
and cosmology of neutrino mass is covered in recent \cite{Dolgov
2002} Their scattering with incoming extra-galactic EHE neutrinos
create relativistic Z whose nucleon decay in cascades could
contribute or dominate the observed UHECR flux at $GZK$ edges.

%%%%%%%%%%%%%%%   Figure 2 Ex 01  %%%%%%%%%%%%%%%%%
\begin{figure}
 \includegraphics[height=8.5cm]{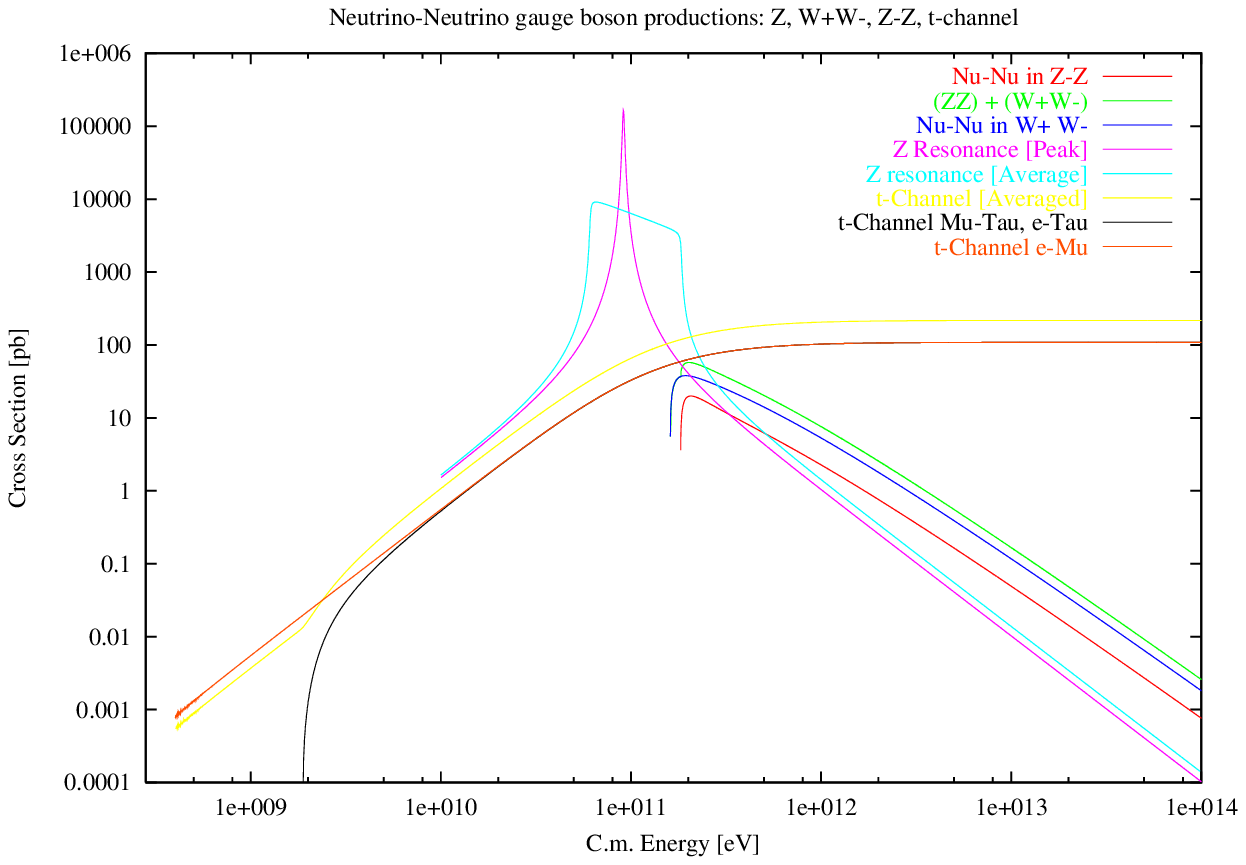}
\centering \caption[h]{The  $\nu \bar{\nu} \rightarrow Z,W^+
W^-,ZZ,T$-channel,  cross sections as a function of the center of
mass energy in $\nu \nu$.  These cross-sections are estimated
also in average (Z) as well for each possible t-channel lepton
pairs. The averaged t-channel averaged the multiplicity of
flavours pairs ${\nu}_{i}$, $ \bar{\nu}_{j}$ respect to neutrino
pair annihilations into Z neutral boson. The Z-WW-ZZ Showering
has to be  boosted by Lorentz transform to show their behaviour
at laboratory system.}
\end{figure}
%%%%%%%%%%%%%%%%%%%%%%%%%%%%%%%%%%%%%%%%%%%%%%%%%%%%%%%%%%%%%%%%%%%%

%%%%%%%%%%%%%%%%%%%%%%%%%%%%%%%%%%%%%%%%%%%%%%%%%%%%%%%%%%%%%%%%%%%%%%%%%%%%%%%
\begin{figure}
\centering
%\epsfysize=0.30\textwidth % fix the y-dimension and scales x-dim. to y-dim.
%\hspace{0.05\textwidth}\epsfbox{table.eps} %for centering: act on hspace argument
 \includegraphics[height=5cm]{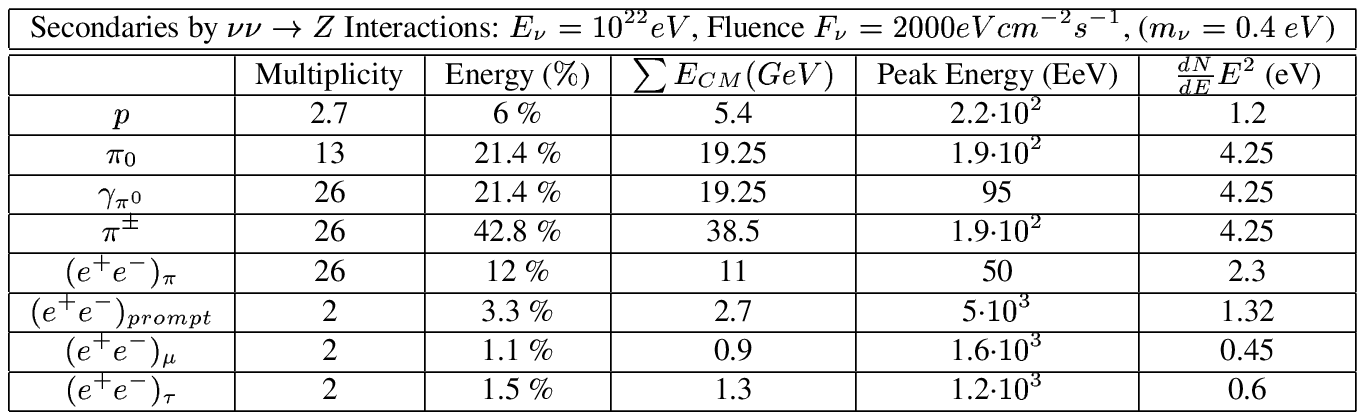}
\caption[h]{Table 1A: The  detailed energy percentage
distribution  into neutrino, protons, neutral and charged pions
and consequent gamma, electron pair particles both from hadronic
and leptonic Z, $WW,ZZ$ channels. We  calculated the
elecro-magnetic contribution due to the t-channel $\nu_i \nu_j$
interactions. We used LEP data for Z decay and considered W decay
roughly in the same way as Z one. We assumed that an average
number of 37 particles is produced during a Z (W) hadronic decay.
The number of prompt pions both charged (18) and neutral (9), in
the hadronic decay is increased by 8 and 4 respectively due to
the decay of $K^0$, $K^{\pm}$, $\rho$, $\omega$, and $\eta$
particles. We assumed that the most energetic neutrinos produced
in the hadronic decay mainly come from charged pion decay. So
their number is roughly three times the number of $\pi$'s. UHE
photons are mainly relics of neutral pions. Most of the $\gamma$
radiation will be degraded around PeV energies by $\gamma \gamma$
pair production with cosmic 2.75 K BBR, or with cosmic radio
background. The electron pairs instead, are mainly relics of
charged pions and will rapidly lose energies into synchrotron
radiation. The contribution of leptonic Z (W) decay is also
considered and calculated in the table above and below.}
\end{figure}
%%%%%%%%%%%%%%%%%%%%%%%%%%%%%%%%%%%%%%%%%%%%%%%%%%%%%%%%%%%%%%%%%%%%%%%%%%
 Neutrino  mass existence related to its flavour mixing has been reinforced
by Super-Kamiokande evidence for atmospheric neutrino anomaly via
$\nu_{\mu} \leftrightarrow \nu_{\tau}$ oscillations and more
compelling evidence from Gallex and SNO solar neutrino mixing
data. Consequently there are at least two main extreme scenario
for hot dark halos: either $\nu_{\mu}\, , \nu_{\tau}$ are both
extremely light ($m_{\nu_{\mu}} \sim m_{\nu_{\tau}} \sim
\sqrt{(\Delta m)^2} \sim 0.05 \, eV$) and therefore energetic and
fast running in hot dark neutrino halo shaped into  wide, smeared
and spread out to local group  sizes,
 or $\nu_{\mu}, \nu_{\tau}$ may share nearly degenerated ($\sim eV$)
 masses. Within the latter fine-tuned neutrino
mass case ($m_{\nu}\sim 0.4  eV$) (see Fig,2), (recently found
possibly hidden in a double beta decay  signal)  the Z peak $\nu
\bar{\nu}_r$ interaction (see Fig.1) (D.Fargion,A.Salis
\cite{Fargion Salis 1997}, D.Fargion, B.Mele, A.Salis
\cite{Fargion Mele Salis 1999}1997-99,Weiler \cite{Weiler1999}
1999, S.Yoshida, G. Sigl, S. Lee \cite{Yoshida et al.1998}1998)
will be the favorite one.

%%%%%%%%%%%%%%%   Figure 3 - 4     %%%%%%%%%%%%%%%%%   FFFFFFFFFFFFFFFFFFFFFFFFFFFFFFFFFFFFFF
%%%%%%%%%%%%%%%   Figure 3 - 4     %%%%%%%%%%%%%%%%%   FFFFFFFFFFFFFFFFFFFFFFFFFFFFFFFFFFFFFF

\begin{figure}
\centering
\includegraphics[width=.49\textwidth]{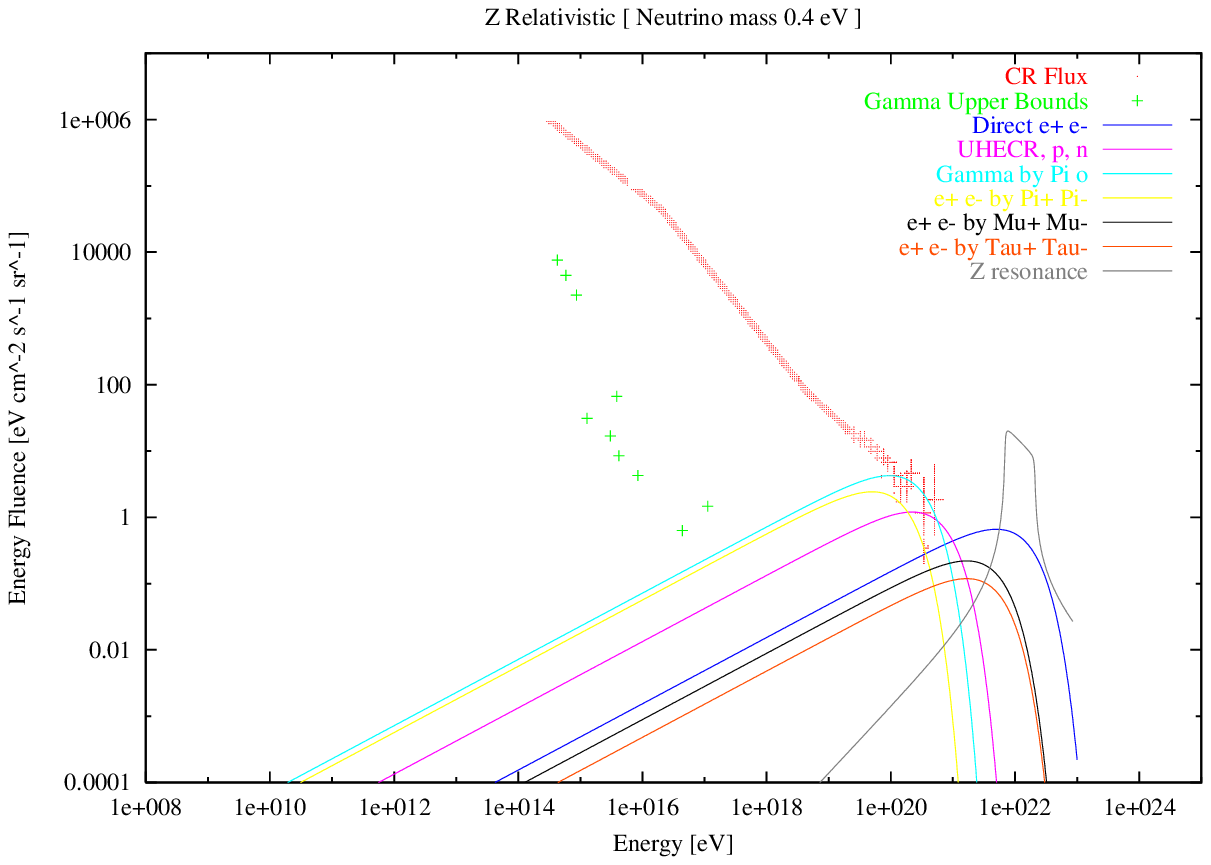}
\includegraphics[width=.49\textwidth]{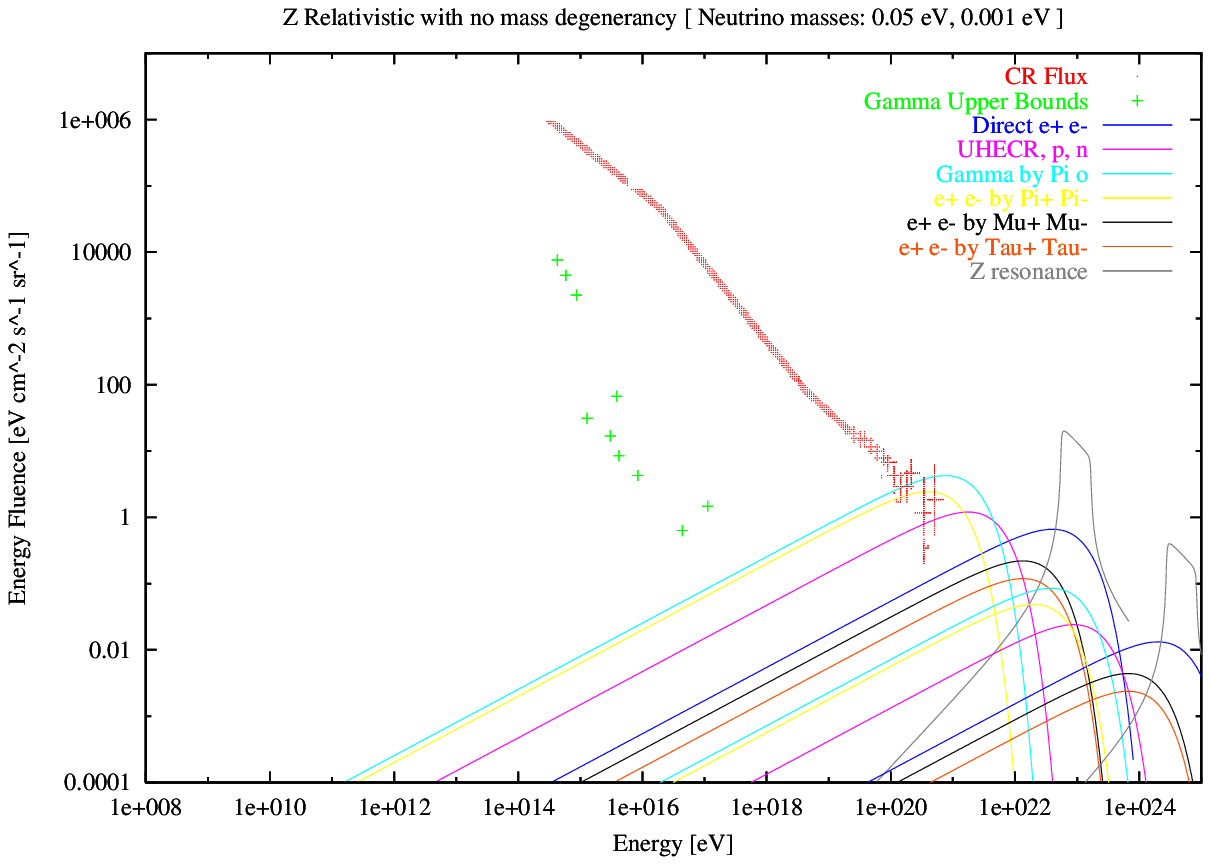}
 \caption[h]{Left: Energy Fluence derived by $\nu \bar{\nu} \rightarrow Z$ and its
 showering into  different channels: direct electron pairs UHECR
 nucleons $n$ $p$ and anti-nucleons, $\gamma$ by $\pi^0$ decay,
  electron pair by $\pi^+ \pi^-$ decay, electron pairs by direct muon and tau decays as labeled in figure.
  The relic neutrino mass has been assumed to be fine tuned to explain GZK UHECR tail:
  $m_{\nu}=0.4 eV$. The Z resonance ghost (the shadows of Z Showering resonance (Fargion 2001) curve),
  derived from Z cross-section in Fig.1, shows the averaged $Z$ resonant cross-section peaked
  at $E_{\nu}=10^{22} eV$. Each channel shower has been normalized following previous table.}
% \end{figure}

%%%%%%%%%%%%%%%%%%%%%BEGIN%%%FIG Twin%%%%%%%%%%%%%%%%%%%%%%%%%%%%%%%%%%%%
%\begin{figure}
\caption[h]{Right: Energy Fluence derived by $\nu \bar{\nu}
\rightarrow Z$ and its showering into  different channels  as
above.  In the present extreme case the relic neutrino masses
have been assumed with wide mass differences
  just compatible both with Super-Kamiokande and relic $2 K^{o}$ Temperature.  The their values have been fine tuned to explain observed GZK- UHECR tail:
   $m_{\nu_1}=0.05eV$ and $m_{\nu_2}=0.001 eV$. A neutrino
   density difference between the two masses  has been
   assumed,considering the lightest $m_{\nu_2}=0.001 eV$ neutrino
   at relativistic regime. The incoming UHE neutrino fluence has been assumed growing
   linearly \cite{Yoshida et al.1998}  with energy. Its value is increased
   by a factor 2 and 20  at  $E_{\nu_1}=8\cdot10^{22} eV$ and $E_{\nu_2}=4\cdot10^{24} eV$
   respect the previous ones Fig.2. The "Z resonance" curve
    shows its averaged $Z$ resonant "ghost" cross-section peaked
  at $E_{\nu_1}=2\cdot10^{23} eV$ and $E_{\nu_2}=4\cdot10^{24} eV$, just
  near Grand Unification energies.   Each channel shower has been normalized in analogy to table 1.}
\end{figure}
%%%%%%%%%%%%%%%%%%%%%%%%%%END%%%%Twin%%%%%%%%%%%%%%%%%%%%%%%%%%%%%

%%%%%%%%%%%%%%%   Figure 3 END   %%%%%%%%%%%%%%%%%

In the second case (for heavier non constrained neutrino mass
($m_{\nu} \gtrsim 2-3 \, eV$)) only a $\nu \bar{\nu}_r
\rightarrow W^+W^-$ D.Fargion,A.Salis \cite{Fargion Salis 1997},
D.Fargion, B.Mele, A.Salis \cite{Fargion Mele Salis 1999}1997-99,
and the additional $\nu \bar{\nu}_r \rightarrow ZZ$ interactions,
(see the cross-section in Fig.1)(Fargion et all. 2001a,b
\cite{Fargion et al.2001a}, \cite{Fargion et al.2001b})
reconsidered here will be the only ones able to solve the GZK
puzzle. Indeed the relic neutrino mass within HDM models in
galactic halo near $m_{\nu}\sim 4 eV$, corresponds to a lower and
$Z$ resonant incoming energy
 ${{E_{\nu} =  {\left(\frac{4eV} {\sqrt{{{m_{\nu}}^2+{p_{\nu}^2}}}} \right)} \cdot
10^{21} \,eV.}}$. This resonant incoming neutrino energy is
showering mainly a small energy fraction into nucleons
($p,\bar{p}, n, \bar{n}$),  (see $Tab.1$ below), at energies
$E_{p}$ quite below: ${{E_{p} =  2.2 {\left(\frac{4eV}
{\sqrt{{{m_{\nu}}^2+{p_{\nu}^2}}}} \right)} \cdot 10^{19} \,eV.}
\nonumber}$. Therefore too heavy ($> 1.5 eV$) neutrino mass are
not fit to  solve GZK by Z-resonance; on the contrary WW,ZZ
showering as well as t-channel showering
   may naturally keep open the solution. In particular the overlapping of both the Z and the WW, ZZ
   channels described in fig.1, for $m_{\nu} \simeq 2.3 eV$ while
   solving the UHECR above GZK they must pile up (by Z-resonance
   peak activity) events at $ 5 \cdot 10^{19} eV$, leading to a bump in
   AGASA data. There is indeed a first marginal evidence of such a UHECR bump
    in AGASA and Yakutsk data that may stand for this interpretation.
     More detailed data are  needed to verify such very exciting  possibility.
\subsection{Relic $\nu_r$ fine-tuned neutrino mass and the Z Showering-Knee }
      Similar result regarding the fine tuned relic mass
    at $0.4 eV$ and $2.3 eV$ (however ignoring the WW ZZ and
    t-channels and invoking very hard UHE neutrino spectra) have been
    independently reported recently (Fodor, Katz, Ringwald \cite{Fodor et al.2001}).
    We notice here (for the first time) and predict that the fine-tuned $\simeq 0.4 eV$ mass
    (following atmospheric and solar mass splitting) being a nearly
    degenerated mass value, must induce, just above $2.2\cdot 10^{20} eV$,
     the UHECR spectra into a very sharp Z-Showering Knee,(or cut-off),
     (see Fig.3 above, UHECR p,n, pink curve);
     this Z-Knee Cut-off  might be soon observable (or not)
      just beyond  (HIRES,AGASA,AUGER) detection corner. The
      lighter the mass the higher energy is the Z-Knee cut-off.
    One of the weak point of Z-Shower model is related to  the relic $\nu$
    number density clustering  that define the probability of the Z Showering.
Indeed let us briefly remind the role of relic $\nu$ mass and
their velocity spread within Fermi-Dirac maximal allowed number
density:$n_{\nu_{i}}=1.9\cdot10^{3}\left(\frac{n_{\nu_{cosmic}}}{54cm^{-3}}\right)
\left( \frac{m_{i}}{0.1eV}\right)^{3}\left(\frac{v_{\nu_{i}}}
{2\cdot10^{3}\frac{Km}{s}}\right)^{3}$. This formula imply that
neutrino number density contrast is bounded by the cube of the
inverse of neutrino mass and by the cube of the local Group
velocity spread.
     The $\nu$ number density problem may be faced either
     with heavier $\nu$ mass or (for  lightest masses)
    with extreme hard UHE neutrino fluxes \cite{Fodor et al.2001}
    or finally with extra relic neutrino degenerancy   \cite{Gelmini et al. 2000}.
   The first case is somehow exaggerating the incoming UHE  $\nu$
   flux, hidden  just below known bounds by Goldstone experiment on ZeV neutrinos.
   The expected $\gamma$ pollution of such huge Z-Showering fluxes
     are also bounded by EGRET data; cosmic degenerancy (chemical
     potential) is increasing only by few unity the target
     density. However (Fargion et all 2001a,2001b) there may  exist,
      well within or beyond Standard Cosmology,
      a relevant relic energetic neutrino ($Mev$)injection due to stellar,
      Super-Novae, AGNs,BLacs,Gamma Ray Bursts, Soft Gamma Repeaters,
      Black Holes or mini-BH  past activities at large redshift ($z\geq 10^6$) ,
      presently  red-shifted   into a  eV $\nu$ spectra,
     piling into a dense relativistic relic neutrino grey-body  spectra.
     In this windy relativistic (or ultra relativistic) neutrino  cosmology,(eventually
      leading to a neutrino radiation dominated Universe), the
      halo size (to be considered) is nearly coincident with
       the GZK one  for UHECR nucleons and gammas($\sim 20 Mpcs$).
       Therefore, while the isotropic UHECR behaviour
      is guaranteed, a puzzle related to the non observed uniform
      spectra  distribution seem to persist. Nevertheless the same
      UHE neutrino-relic neutrino scattering  cross-sections (figure 1) \textit{do not} follow a flat
      spectra, (as well as any convolutions with hypothetical $\nu$ grey  body spectra).
      This may explain why we may leave in a
      homogeneous  relic relativistic neutrino component at eVs energies as
      well as why we do observe a  non-uniform UHECR spectra (reflecting non-homogeneous Z-WW-ZZ channels).

      % This case is similar to the
      % case of a very light neutrino mass much below $0.1$ eV.
%As we noticed above, relic neutrino mass above a few eVs in  HDM
%halo \textit{are not} consistent with naive Z peak; higher
%energies interactions ruled by WW,(D.Fargion, B.Mele et A.Salis
%1999; K.Enqvist et all. 1989) ZZ cross-sections (Fargion 2001) may
%nevertheless solve the GZK cut-off. In this regime there will be
%also possible to produce by virtual W exchange, t-channel, $UHE$
%lepton pairs, by $\nu_i \bar{\nu}_j\rightarrow l_i\bar{l}_j$,
%leading to additional electro-magnetic showers injection.\\

Important and often underestimated signal will produce UHE
electrons pairs( by Z decay) whose final trace are TeVs  photons
able to break the IR-TeV cosmic cut-off. The hadronic tail of the
Z or $W^+ W^-$ cascade maybe the source of final  nucleons
$p,\bar{p}, n, \bar{n}$ able to explain UHECR events. The same
$\nu \bar{\nu}_r$ interactions are source of Z and W that decay
in rich shower ramification \cite{Fargion et al.2001b}

%\section{UHECR  channels in Z showers}
% Although protons (or anti-protons)
%  are the most popular and favorite candidate in
%order to explain the highest energy air shower observed, one
%doesn't have to neglect the signature of final neutron and
%anti-neutrons as well as electrons and photons. Indeed the UHECR
%neutrons are produced in Z-WW showering at nearly same rate as
%the charged nucleons. Above GZK cut-off energies UHE $n$,$
%\bar{n}$, share a life lenght comparable with the Hot Galactic
%Dark Neutrino Halo. Therefore they may be an important component
%in UHECRs. Moreover prompt UHE electron (positron) interactions
%with the galactic or extra-galactic magnetic field or soft
%radiative backgrounds may lead to gamma cascades and from PeVs to
%TeVs energies.\\
Gamma photons at energies $E_{\gamma} \simeq 10^{20}$ - $10^{19}
\,eV$, secondary of Z-Showering cascades, may freely propagate
through galactic or local halo scales (hundreds of kpc to few
Mpc) and could also contribute to the extreme edges of cosmic ray
spectrum  and clustering ( see:\cite{Yoshida et al.1998},
\cite{Fargion et al.2001a}, \cite{Fargion et al.2001b}).  The
ratio of the final energy flux of nucleons near the Z peak
resonance, $\Phi_p$ over the corresponding electro-magnetic
energy flux $\Phi_{em}$ ratio is, as in tab.1 $e^+ e^-,\gamma$
entrance, nearly $\sim \frac{1}{8}$.  Moreover if one considers
at higher $E_{\nu}$ energies, the opening of WW, ZZ channels and
the six pairs $\nu_e \bar{\nu_{\mu}}$, \, $\nu_{\mu}
\bar{\nu_{\tau}}$, \, $\nu_e \bar{\nu_{\tau}}$ (and their
anti-particle pairs) t-channel interactions leading to  highest
energy leptons, with no nucleonic relics (as $p, \bar{p}$), this
additional injection favors the electro-magnetic flux $\Phi_{em}$
over the corresponding nuclear one $\Phi_p$ by a factor $\sim
1.6$ leading to $\frac{\Phi_p}{\Phi_{em}} \sim \frac{1}{13}$.
This ratio is valid at $WW,ZZ$ masses because the overall cross
section variability is energy dependent. At center of mass
energies above these values, the $\frac{\Phi_p}{\Phi_{em}}$
decreases more because the dominant role of t-channel (Fig1). We
focus here  on Z, and WW,ZZ channels showering in hadrons for GZK
events. The important role of UHE electron showering into TeV
radiation is discussed below.
%\subsection{UHE $\nu$ - $\nu_{relic}$ Cross Sections }
 % There is an upper bound density clustering for
%very light Dirac fermions due to the maximal Fermi degenerancy
%whose adimensional density contrast is $\delta\rho \propto
%m_{\nu}^3$,  while one finds (Fargion 1983)  that  the neutrino
%free-streaming halo grows only as $\propto m_{\nu}^{-1}$.
%Therefore the overall interaction probability grows $ \propto
%m_{\nu}^{2} $, favoring heavier non relativistic (eVs) neutrino
%masses. In this frame above few eV neutrino masses only WW-ZZ
%channel are operative. Nevertheless the same lightest relic
%neutrinos may share higher Local Group velocities (thousands
%$\frac{Km}{s}$) or even nearly relativistic speeds and it may
%therefore compensate the common density bound:

%\begin{figure}
% \includegraphics[height=8.5cm]{nu12}
%\caption[h]{Energy
%Fluence derived by $\nu \bar{\nu} \rightarrow Z$ and its
%showering into different channels as in previous Figure 2: direct electron pairs UHECR nucleons $n$ $p$, $\gamma$ by $\pi^0$ decay,
 % electron pair by $\pi^+ \pi^-$ decay, electron pairs by direct muon and tau decays as labeled in figure.
 % In the present case the relic neutrino mass has been assumed to be fine tuned to explain GZK UHECR tail:
 % $m_{\nu}=1.2 eV$ with the same UHE incoming neutrino fluence of previous figure. The Z resonance curve shows the averaged $Z$ resonant cross-section peaked
 % at $E_{\nu}=3.33\cdot10^{21} eV$.Each channel shower has been normalized in analogy to table 1B.}
%\end{figure}
%%%%%%%%%%%%%%%   Figure 4 END   %%%%%%%%%%%%%%%%%
The minimal neutrino mass  (near 0.1 eV) comparable with present
Super-Kamiokande atmospheric neutrino mass splitting  are leading
to an exciting scenario where (more than one) non degenerated
Z-resonances occur (Fargion et all.2001a,b \cite{Fargion et
al.2001a}\cite{Fargion et al.2001b}). These scenario are
summarized in (Fig. 2, in Ref.\cite{Fargion et al.2001a} and Ref.
\cite{Fargion et al.2001b}), (for nominal example
$m_{\nu_{\tau}}$ = 0.1 eV; $m_{\nu_{\mu}}$ = 0.05 eV). The twin
neutrino mass inject a corresponding twin bump at highest energy.
Another limiting case of interest takes place when the light
neutrino masses are both extreme, nearly at atmospheric (SK,K2K)
and solar (SNO) neutrino masses. This case is described in
following Fig.3( keeping care of the lower number density for
lightest neutrino mass). The relic neutrino masses are assumed
$m_{\nu_{\tau}}$ = 0.05 eV; $m_{\nu_{\mu}}$ = 0.001 eV.
The  neutrino mass play a role also in defining its Hot Dark Halo
size and our peculiar position in such HDM halo. Indeed for a
heavy $\geq 2 eV$ mass case HDM neutrino halo are mainly galactic
and/or local, reflecting an isotropic or a diffused amplification
toward nearby $M31$ HDM halo. In the lighter case the HDM should
include the Local Cluster up to Virgo . To each size corresponds
also a different role of UHECR arrival time. The larger the HDM
size the longer the UHECR random-walk travel time and the longer
the lag between doublets or triplets.
% FORSE FALSO...:The larger the neutrino mass,the smaller is the neutrino halo
%(quadratic inverse of neutrino mass) the
%earlier the UHE neutron secondaries by Z shower (linear in m_nu mass)
%will play a role:
%indeed at $E_{n}= 10^{20}eV$ UHE neutron are flying a Mpc and
%their directional arrival (or their late decayed proton arrival)
%are more on-line toward the source. This may explain the high self
%collimation and auto-correlation of UHECR discovered very
%recently (Tinyakov et all. 2001).
 The UHE neutrons by Z-Showering fits naturally the harder spectra observed in clustered events in
AGASA ( \cite{Takeda et al.2001} 2001). The same UHECR neutrons
may explain the quite short (2-3 years)\cite{Takeda et al.2001}
lapse of time observed in AGASA doublets.
%Indeed the most conservative scenario where UHECR
%are just primary proton from nearby sources at GZK distances
%(tens of Mpcs) are no longer acceptable either because the
%absence of such nearby sources candidate and because of the
%observed stringent UHECR clustering ($2^o - 2.5^o$) (Takeda et
%all. 2001) in arrival direction, as well as because of the short
%($\sim3$ years) characteristic time lag between clustered events.
%\subsection{The apparent Tinyakov et Tkachev-Glushkov  Paradox }
The same role of UHE neutron secondaries from Z showering in HDM
halo may also solve an emerging puzzle: the  correlations of
arrival directions of UHECRs found recently (Glushkov et
all.\cite{Glushkov et al.2001} 2001) in Yakutsk data at energy $E=
8\cdot 10^{18} eV$ toward the Super Galactic Plane are to be
compared with the compelling evidence of UHECRs events ($E=
3\cdot 10^{19} eV$ above GZK) clustering toward well defined BL
Lacs at cosmic distances (redshift $z> 0.1-0.2$) (Tinyakov et
Tkachev \cite{Tinyakov-Tkachev2001} 2001; Gorbunov et al.
\cite{Gorbunov et al.2002} 2002). The question arise: where is the
real UHECR sources location? At Super-galactic disk ($50$ Mpcs
wide, within GZK range) or at cosmic ($\geq 300Mpcs$) edges? Of
course both results (or just one of them) maybe a statistical
fluctuation. But both studies seem statistically significant
(4.6-5 sigma) and they seem in some obvious disagreement. There
may be  a possibility for $two$ new categories of UHECR sources.
But it seem quite unnatural any propagation of direct nucleons
because the UHECR from most distant BLac sources are the harder.
However our Z-Showering scenario offer a common simultaneous
solution: (1) The Relic Neutrino Masses define different
Hierarchical Dark Halos and arrival direction correlated to Hot
(Anisotropic) Relic Neutrino Halos. The real sources are at
(isotropic) cosmic edges \cite{Gorbunov et al.2002}   but their
crossing along a wider anisotropic relic neutrino cloud enhance
the interaction probability in the Super Galactic Plane
\cite{Gorbunov et al.2002}. (2) The nearest SG sources are weaker
while the collimated BL Lacs are harder: anyway  both sources need
a Neutrino Halo to induce the Z-Showering UHECRs.
\subsection{ The TeV Tails from UHE Z-WW-ZZ Showering into secondary electrons }
 As it is shown in Table 1 and Figures above, the electron
(positron) energies by $\pi^{\pm}$ decays is around $E_e \sim 2
\cdot 10^{19} \, eV$ for an initial $E_Z \sim 10^{22} \, eV $ (
and $E_{\nu} \sim 10^{22} \, eV $). Such electron pairs while not
radiating efficiently in low extra-galactic magnetic fields they
will be interacting with the galactic magnetic field ($B_G \simeq
10^{-6} \,G $)  leading to direct TeV photons:
%\begin{displaymath}
 $ E_{\gamma}^{sync} \sim \gamma^2 \left( \frac{eB}{2\pi m_e } \right)
    \sim 27.2 \left( \frac{E_e}{2 \cdot10^{19}
  \,eV} \right)^2 \left( \frac{m_{\nu}}{0.4 \, eV} \right)^{-2} \left( \frac{B}{\mu G} \right)\,TeV.
$
%\end{equation}
The same UHE electrons will radiate less efficiently with extra-
galactic magnetic field ($B_G \simeq 10^{-9} \,G $)  leading also
to direct peak $27.2$ GeV  photons.
   The spectrum of these photons is characterized by a power of law $dN
/dE dT \sim E^{-(\alpha + 1)/2} \sim E^{-1.25}$ where $\alpha$ is
the power law of the electron spectrum, and it is showed in
Figures above. As regards the prompt electrons at higher energy
($E_e \simeq 10^{21}\, eV$), in particular in the t-channels,
their interactions with the extra-galactic field first and
galactic magnetic fields later is source of another kind of
synchrotron emission around tens of PeV energies (Fargion et all
\cite{Fargion et al.2001b} 2001).
%$E^{sync}_{\gamma}$: % E^{sync}_{\gamma}
%$ \sim 6.8 \cdot 10^{13} \left( \frac{E_e}{10^{21}\,eV} \right)^2
%  \left( \frac{m_{\nu}}{0.4 \, eV} \right)^{-2} \left(
%\frac{B}{nG}
%  \right) \, eV.\sim
%  6.8 \cdot 10^{16} \left( \frac{E_e}{10^{21}\,eV} \right)^2
%  \left( \frac{m_{\nu}}{0.4 \, eV} \right)^{-2} \left( \frac{B}{\mu G}
%  \right) \, eV

%\end{equation}
%The corresponding energy loss length instead is (O.E.Kalashev,
%V.A.Kuzmin, D.V.Semikoz 2000)
%\begin{equation}\label{3}
%$\left( \frac{1}{E} \frac{dE}{dt} \right)^{-1} = 3.8 \times \left(
%\frac{E}{10^{21}} \right)^{-1} \left( \frac{B}{10^{-9} G}
%\right)^{-2} \, kpc.
%\end{equation}
 %For the first case the interaction lenght is few Kpcs while in
%the second one in few days light flight. Again one has the same
%power law characteristic of a synchrotron spectrum with index
%$E^{-(\alpha + 1 / 2)} \sim E^{-1.25}$.
 Gammas at $10^{16} \div 10^{17}$ eV scatters onto
low-energy photons from isotropic cosmic background ($\gamma + BBR
\rightarrow e^+ e^-$) converting their energy in electron pair.
 %The expression of the pair production cross-section is:
%\begin{equation}
%$ \sigma (s) = \frac{1}{2} \pi r_0^2 (1 - v^2) [
%(3 - v^4) \ln \frac{1 + v}{1 - v} - 2 v (2 - v^2) ]
%$
%\end{equation}
%where $v = (1 - 4m_e^2 / s)^{1/2}$,  $s = 2 E_{\gamma} \epsilon (1
%- \cos \theta)$ is the square energy in the center of mass frame,
%$\epsilon$ is the target photon energy, $r_0 $  is the classic
%electron radius, with a peak cross section value at
%$frac{4}{137}\times \frac{3}{8\pi} \sigma_T \ln 183 = 1.2 \times
%10^{-26} \,cm^2 $.
 Because the corresponding attenuation length
due to the interactions with the microwave background is around
ten kpc, the extension of the halo plays a fundamental role in
order to make this mechanism efficient or not. As is shown in
figures above the contribution to tens of PeV gamma signals by Z
(or W) hadronic decay, could be compatible with actual
experimental limits fixed by CASA-MIA detector on such a range of
energies. Considering a halo extension $l_{halo} \gtrsim 100
kpc$, the secondary electron pair creation becomes efficient,
leading to a suppression of the tens of PeV signal. So electrons
at $E_e \sim 3.5 \cdot 10^{16} \,eV$ loose again energy through
additional synchrotron radiation with maximum $E_{\gamma}^{sync}$
around
%\begin{equation}\label{3b}
 $ \sim 79 \left( \frac{E_e}{10^{21}
  \,eV} \right)^4 \left( \frac{m_{\nu}}{0.4 \, eV} \right)^{-4}
   \left( \frac{B}{\mu G} \right)^3 \, MeV.
$
%\end{equation}
Anyway this signal is not able to pollute sensibly the MeV-GeV
while its  relevance is striking by a pile up signal at TeVs.
%%%%%%%%%%%%%%%%%%%%%%%%%%%%%%%%%%%%%%%%%%%%%%%%%%%%%%%%%%%%%%%%%%%%%%%%%%%%%%%5
In this frame let us remind that Gamma rays with energies up to
$20$ TeV have been observed by terrestrial detector only by nearby
sources like Mrk 501 (z $= 0.033$) or very recently by MrK $421$
(z $= 0.031$). More recent evidences of tens TeVs from  (three
times more) distant blazar 1ES1426+428 (z $= 0.129$) make even
more dramatic the IR-TeV cut-off. This is puzzling because the
extra-galactic TeV spectrum should be, in principle,
significantly suppressed by the $\gamma$-rays interactions with
the extra-galactic Infrared background, leading to electron pair
production and TeVs cut-off. The recent calibration and
determination of the infrared background by DIRBE and FIRAS on
COBE have inferred severe constrains on TeV propagation. Indeed,
as noticed by Kifune (\cite{Kifune 1999} 1997), and Protheroe and
Meyer \cite{Protheroe and Meyer 2000}  we may face a severe
infrared background - TeV gamma ray crisis. This crisis imply a
distance cut-off, incidentally, comparable to the GZK one. So our
present Z-Showering scenario (\cite{Fargion et al.2001b}) may
easily solve also the IR-TeV cut off. Let us remind also an
additional evidence for IR-TeV cut-off is related to the possible
discover of tens of TeV counterparts of BATSE GRB970417, observed
by Milagrito(R. Atkins et all \cite{Atkins et.al.2000} 2000),
being most GRBs very possibly at cosmic edges, at distances well
above the IR-TeV cut-off ones.
% In this scenario it
%is also important to remind the possibilities that the Fly's Eye
%event has been correlated to TeV pile up events in HEGRA (Horns
%et al. 1999). The very recent report (private communication 2001)
%of the absence of the signal few years  later at HEGRA may be
%still consistent with a limited UHE TeV tail activity.
 To solve the IR-TeV cut-off one may alternatively invoke unbelievable extreme hard intrinsic
spectra or exotic explanation as gamma ray superposition of
photons or finally sacrilegious  Lorentz invariance violation.
%%%%%%%%%%%%%%%%%%%%%%%%%%%%%%%%%%%%%%%%%%%%%%%%%%%%%%%%%%%%%%%%%
%%%%%%%%%%%%%%%   Figure 7       %%%%%%%%%%%%%%%%%   FFFFFFFFFFFFFFFFFFFFFFFFFFFFFFFFFFFFFF
\begin{figure}
\includegraphics[width=.49\textwidth]{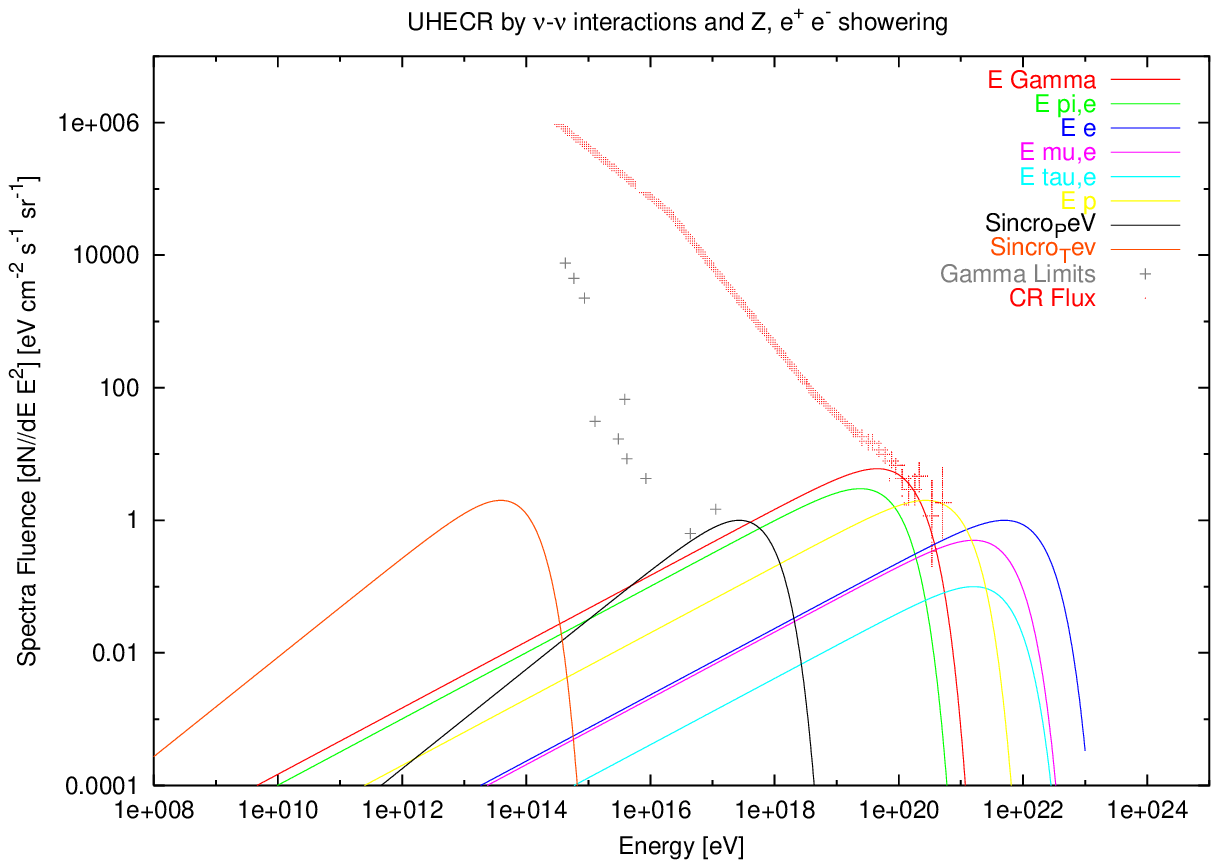}
\includegraphics[width=.49\textwidth]{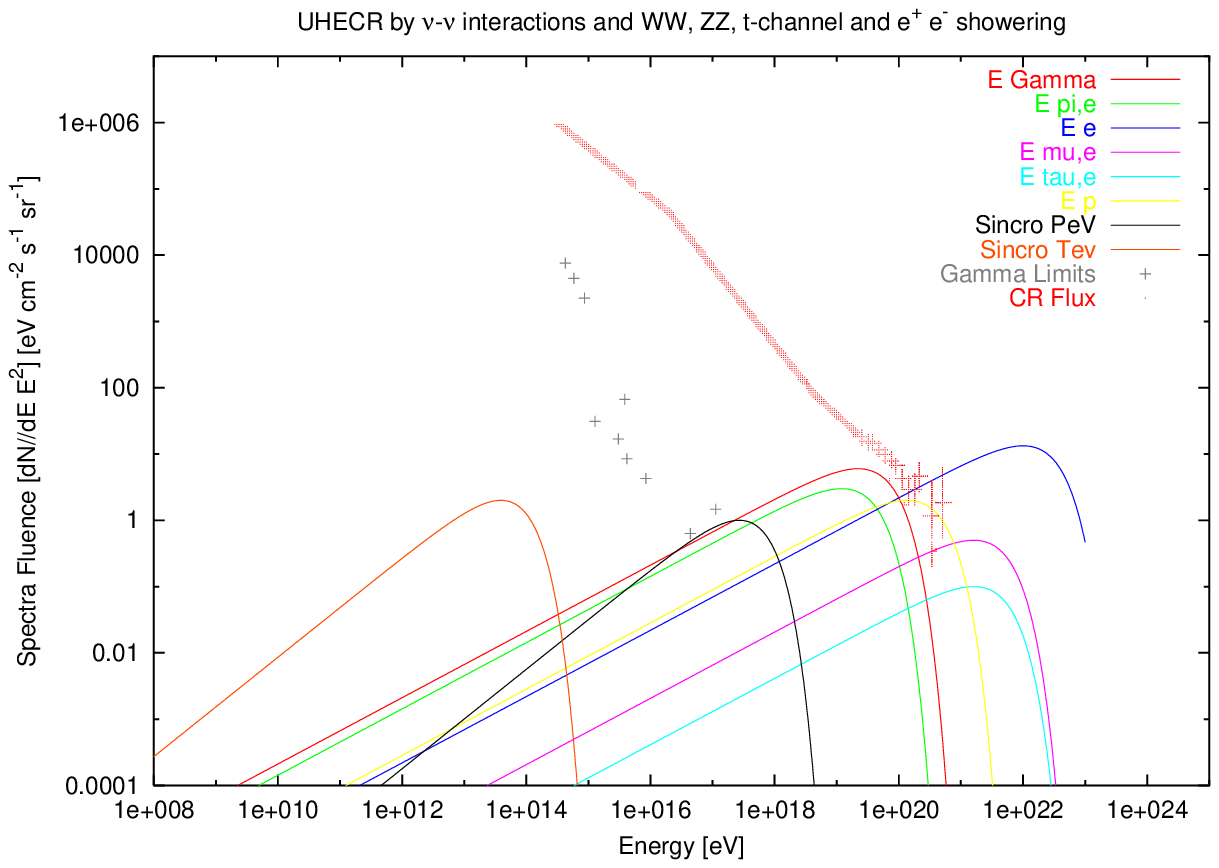}
 \caption[h]{Left:Energy
fluence by Z showering as above for $m_{\nu} = 0.4 eV$ and
$E_{\nu} = 10^{22} eV $and the consequent $e^+ e^-$ synchrotron
radiation }
%\end{figure}

%%%%%%%%%%%%%%%   Figure 7END   %%%%%%%%%%%%%%%%%   FFFFFFFFFFFFFFFFFFFFFFFFFFFFFFFFFFFFFF

%%%%%%%%%%%%%%%   Figure 8       %%%%%%%%%%%%%%%%%   FFFFFFFFFFFFFFFFFFFFFFFFFFFFFFFFFFFFFF
%\begin{figure}
\caption[h]{Right: Energy fluence by WW, ZZ, t-channel showering
as in fig.1, for $m_{\nu} = 0.4 eV$ and $E_{\nu} = 2 \cdot 10^{22}
eV $, and the consequent $e^+ e^-$ synchrotron radiation. The
lower energy Z showering is not included to make spectra more
understandable}
\end{figure}
%%%%%%%%%%%%%%%   Figure 8END   %%%%%%%%%%%%%%%%%
%%%%%%%%%%%%%%%%%%%%%%%%%%%%%%%%%%%%%%%%%%%%%%%%%%%%%%%
%%%%%%%%%%%RATES 2%%%%%%%%%%%%%%%%%%%%%%%%%
%

\def\lesssim{\mathrel{\hbox{\rlap{\hbox{\lower4pt\hbox{$\sim$}}}\hbox{$<$}}}}
\def\gtrsim{\mathrel{\hbox{\rlap{\hbox{\lower4pt\hbox{$\sim$}}}\hbox{$>$}}}}

To conclude the puzzle one finally needs to
scrutiny the UHE $\nu$ astronomy and to test the GZK solution
within Z-Showering Models by any independent search on Earth for
such UHE neutrinos traces above PeVs reaching either EeVs-ZeVs
extreme energies.
 \subsection{UHE $\nu$ Astronomy by $\tau$ Air-Shower}
 Recently Fargion et al.1999,\cite{Fargion et al.1999},Fargion 2002\cite{Fargion 2000-2002}
 proposed a new competitive UHE $\nu$ detection based on ultra high energy
$\nu_{\tau}$ interaction in matter and its consequent secondary
$\tau$ decay in flight while escaping from the rock (Mountain
Chains) or water (Sea)  in air leading to Upward or Horizontal
$\tau$ Air-Showers (UPTAUs and HORTAUs),Fargion $
2001a$,\cite{Fargion 2001a},Fargion $2001b$,\cite{Fargion 2001b}.
In a pictorial way one may compare the UPTAUs and HORTAUs as the
double bang processes expected in $km^3$ ice-water volumes
Learned Pakvasa 1995,\cite{Learned Pakvasa 1995}: the double bang
is due first to the UHE $\nu_{\tau}$ interaction in matter and
secondly by its consequent $\tau$ decay in flight. Here we
consider  a (hidden) UHE $\nu$-N Bang $in$ (the rock-water within
a mountain or the Earth Crust) and a $\tau$ bang $out$ in air,
whose shower is better observable at high altitudes. The main
power of the UPTAUs and HORTAUs detection is the huge
amplification of the UHE neutrino signal, which may deliver
almost all its energy in numerous secondaries traces (Cherenkov
lights, gamma, X photons, electron pairs, collimated muon
bundles). Indeed the multiplicity in $\tau$ Air-showers secondary
particles, $N_{opt} \simeq 10^{12} (E_{\tau} / PeV)$, $
N_{\gamma} (< E_{\gamma} > \sim  10 \, MeV ) \simeq 10^8
(E_{\tau} / PeV) $ , $N_{e^- e^+} \simeq 2 \cdot 10^7
(E_{\tau}/PeV) $ , $N_{\mu} \simeq 3 \cdot 10^5
(E_{\tau}/PeV)^{0.85}$ makes easy the UPTAUs-HORTAUs discover.
 These HORTAUs, also named Skimming neutrinos \cite{Feng et al. 2002}, maybe also
originated on front of mountain chains \cite{Fargion et
al.1999},\cite{Fargion 2000-2002}, \cite{Hou Huang 2002} either by
$\nu_{\tau}N$, $ \bar\nu_{\tau}N$ interactions as well as by $
\bar\nu_{e} e \rightarrow W^{-} \rightarrow \bar\nu_{\tau} \tau$.
This new UHE $\nu_{\tau}$ detection is mainly based on the
oscillated UHE neutrino $\nu_{\tau}$ originated by more common
astrophysical $\nu_{\mu}$, secondaries of pion-muon decay at
PeVs-EeVs-GZK energies. These oscillations are guaranteed  by
Super Kamiokande evidences for flavour mixing within GeVs
atmospheric neutrino data as well as by most solid and recent
evidences of complete solar neutrino mixing observed by SNO
detector. HORTAUs from mountain chains must nevertheless occur,
even for no flavour mixing, as being inevitable $\bar\nu_{e}$
secondaries of common pion-muon decay chains ($\pi^{-}\rightarrow
\mu^{-}+\bar\nu_{\mu}\rightarrow e^{-}+\bar\nu_{e}$) near the
astrophysical sources at Pevs energies. They are mostly absorbed
by the Earth and are only rarely arising as UPTAUS. Their Glashow
resonant interaction allow them to be observed as HORTAUs only
within a very narrow and nearby crown edges at horizons (not to be
discussed here). At wider energies windows ($10^{14}eV-
10^{20}eV$) only neutrino $\nu_{\tau}$, $\bar{\nu}_{\tau}$ play a
key role in UPTAUS and HORTAUS. These Showers might be easily
detectable looking downward the Earth's surface from mountains,
planes, balloons or satellites observer. Here the Earth itself
acts as a "big mountain" or a wide beam dump target. The present
upward $\tau$ at horizons should not be confused with an
independent and well known, complementary (but rarer) Horizontal
Tau Air-shower originated inside the same terrestrial atmosphere:
we shall referee to it as the Atmospheric Horizontal Tau
Air-Shower. The same UPTAUS have a less competitive upward
showering due to $\nu_{e}$ $\bar\nu_{e}$ interactions with
atmosphere, showering in thin upward air layers \cite{Berezinsky
1990}: let us label this atmospheric Upward Tau as A-UPTAUs  and
consider its presence  as a very small additional contribute,
because rock is more than $3000$ times denser than air (see the
ratio in last column in final Table). Therefore at different
heights we need to estimate the UPTAUS and HORTAUs event rate
occurring along the thin terrestrial crust below the observer,
keeping care of their correlated  variables.
\section{THE  UPTAUs-HORTAUs Skin Earth Crowns }
The $\tau$ airshowers are observable at different height $h_{1}$
 leading to different underneath observable terrestrial
areas and crust volumes. HORTAUs in deep valley are also relate
to the peculiar geographical morphology and composition
\cite{Fargion 2000-2002}, as discussed below \cite{Fargion 2002}.
We remind in this case the very important role of UHE  $
\bar\nu_{e}e \rightarrow W^{-}\rightarrow \bar\nu_{\tau}\tau^{-}
$ channels which may be well observable even in absence of any
$\nu_{\tau}$, $ \bar\nu_{\tau}$ UHE sources or any neutrino
flavour mixing: its Glashow peak resonance make these neutrinos
unable to cross all the Earth across but it may be observable
beyond mountain chain \cite{Fargion 2000-2002}; while testing
$\tau$ air-showers beyond a mountain chain one must keep in  mind
the possible amplification of the signal because of a possible
New TeV Physics (see Fig 9) \cite{Fargion 2000-2002}. In the
following we shall consider in general the main $\nu_{\tau}-N$,$
\bar\nu_{\tau}-N$ nuclear interaction on Earth crust. It should
be kept in mind also that UPTAUs and in particular HORTAUS are
showering at very low densities and their geometrical opening
angle (here assumed at $\theta\sim 1^o$) is not in general
conical (like down-ward showers) but they are more in a thin
fan-like shape (like the observed  $8$ shaped horizontal
Air-Showers). The fan shape is opened by the Terrestrial magnetic
field bending. These UPTAUs-HORTAUs duration time is also much
longer than common down-ward showers because their showering
occurs at much lower air density: from micro (UPTAUS from
mountains) to millisecond (UPTAUs and HORTAUs from satellites)
long flashes. Indeed the GRO observed Terrestrial Gamma Flashes,
possibly correlated with the UPTAUs \cite{Fargion 2000-2002} show
the millisecond duration times. In order to estimate the rate and
the fluence for of UPTAUs and HORTAUs one has to  estimate the
observable mass, facing a complex chain of questions, leading for
each height $h_{1}$, to an effective observable surface and
volume from where UPTAUs and HORTAUs might be originated
\cite{Fargion 2002}. From this effective volume it is easy to
estimate the observable rates, assuming a given incoming UHE
$\nu$ flux model for galactic or extragalactic sources. Here we
shall only refer to the Masses estimate,unrelated to any UHE
$\nu$ flux models. These steps are linking simple terrestrial
spherical geometry and its different geological composition, high
energy neutrino physics and UHE $\tau$ interactions, the same UHE
$\tau$ decay in flight and its air-showering physics at different
quota within terrestrial air density \cite{Fargion 2002}.
Detector physics threshold and background noises, signal rates
have been kept in mind \cite{Fargion 2000-2002}, but they will be
discussed and explained in  forthcoming papers.
%\section{The Skin Crown Earth Volumes for HORTAUs}
Let us therefore define, list and estimate below the sequence of
the key variables whose dependence (shown below or derived in
Appendices) leads to the desired HORTAUs volumes (useful to
estimate the UHE $\nu$ prediction rates) summirized in a last
Table and in Conclusions. Let us now  show the main functions
whose interdependence with the observer altitude lead to estimate
the UPTAUs and HORTAUs equivalent detection Surfaces, Volumes and
Masses \cite{Fargion 2002}.The horizontal distance $d_{h}$
toward the horizons:
 % \begin{displaymath}
%d_{h} =  \sqrt{( R_{\oplus} + h_1)^2 - (R_{\oplus})^2}=
 % \end{displaymath}
 %\begin{equation}
 $d_{h}= 113\sqrt{ \frac{h_1}{km} }\cdot
 \sqrt{1+\frac{h_1}{2R_{\oplus}}}{km}.
  $
The corresponding horizontal edge angle $\theta_{h}$:
 ${\theta_{h} }={\arccos {\frac {R_{\oplus}}{( R_{\oplus} + h_1)}}}\simeq 1^o \sqrt{\frac
 {h_{1}}{km}}.
$%\end{equation}
(Approximations here and below hold for
height$h_{1}\ll{R_{\oplus}}$.)
\begin{figure}
\centering
\begin{minipage}[c] {0.3\textwidth}
\includegraphics[width= 1 \textwidth]{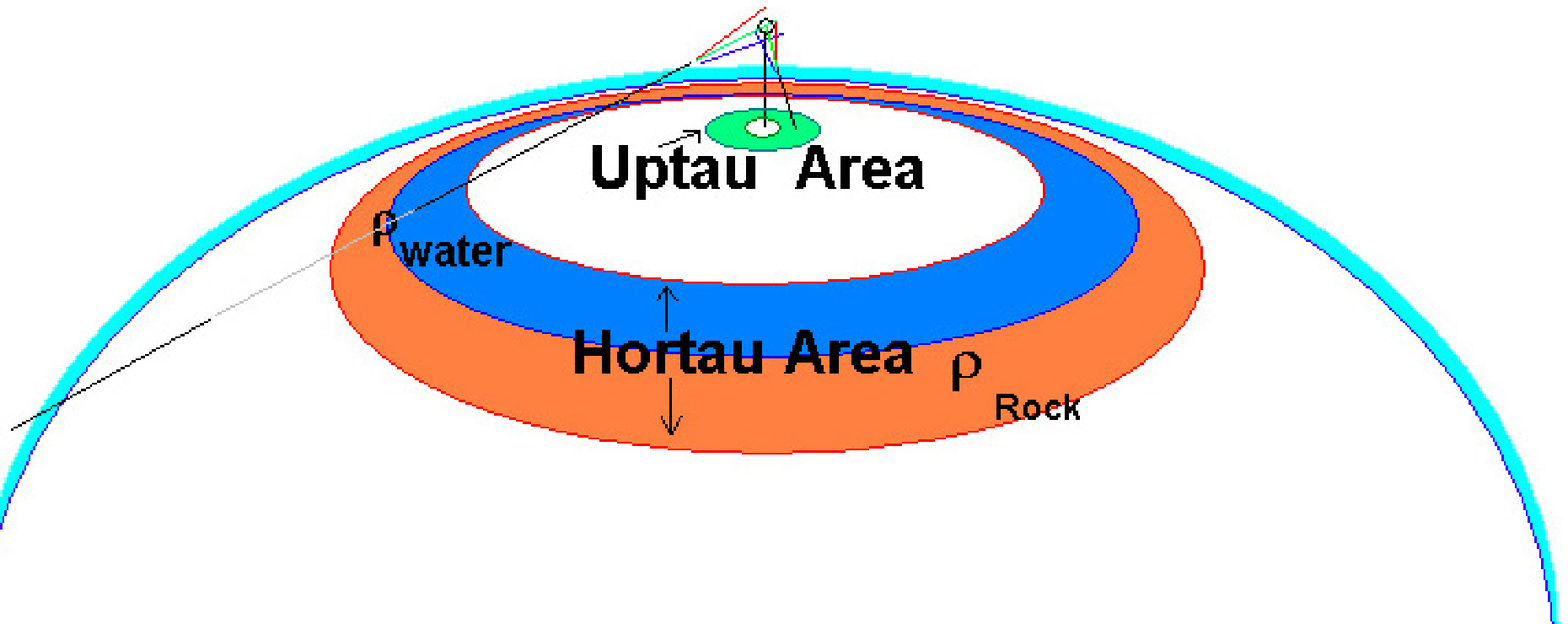}
\end{minipage}%
\begin{minipage}[c] {0.3 \textwidth}
\includegraphics[width= 0.9 \textwidth]{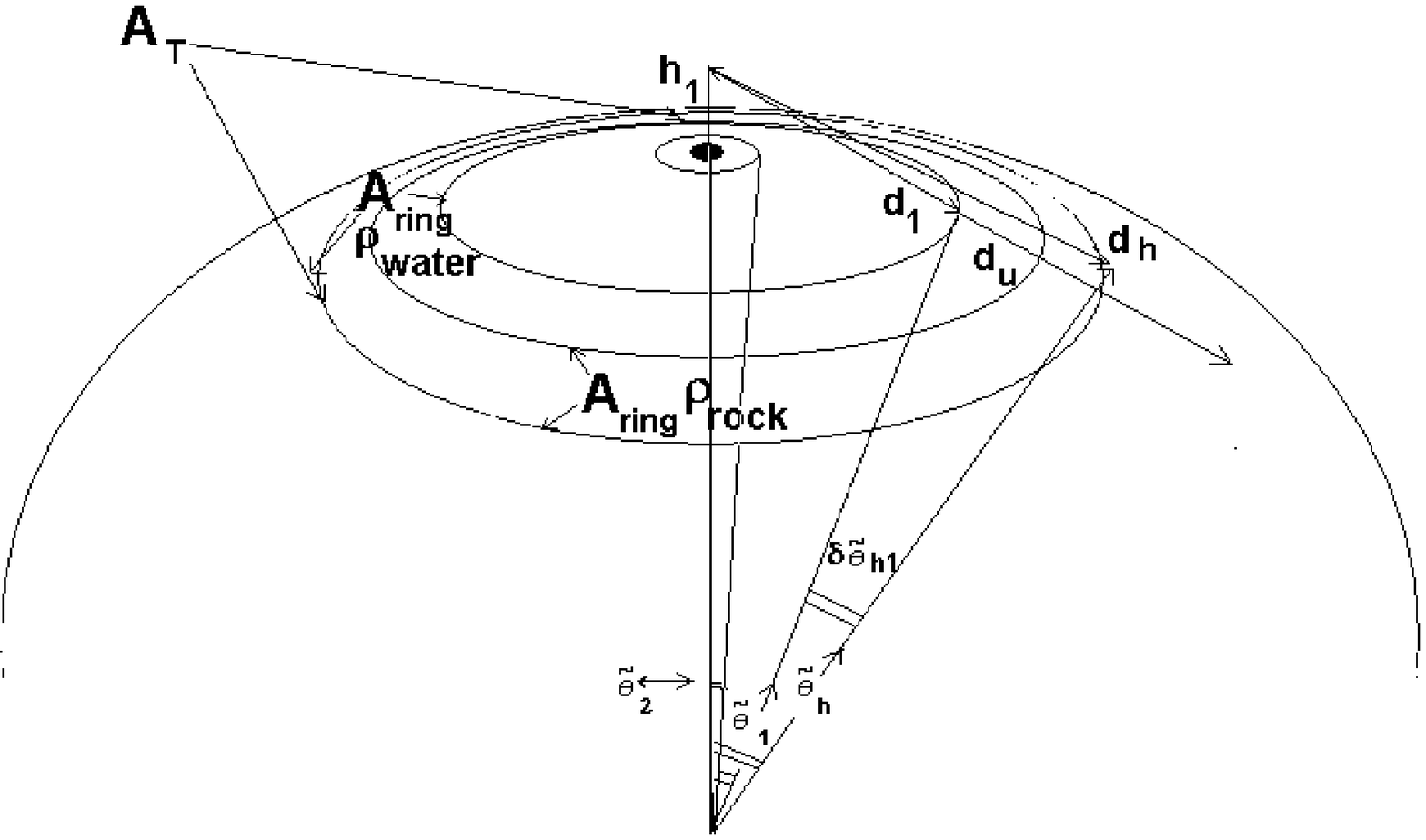}
\end{minipage}
\begin{minipage}[c] {0.3 \textwidth}
\includegraphics[width= 1 \textwidth]{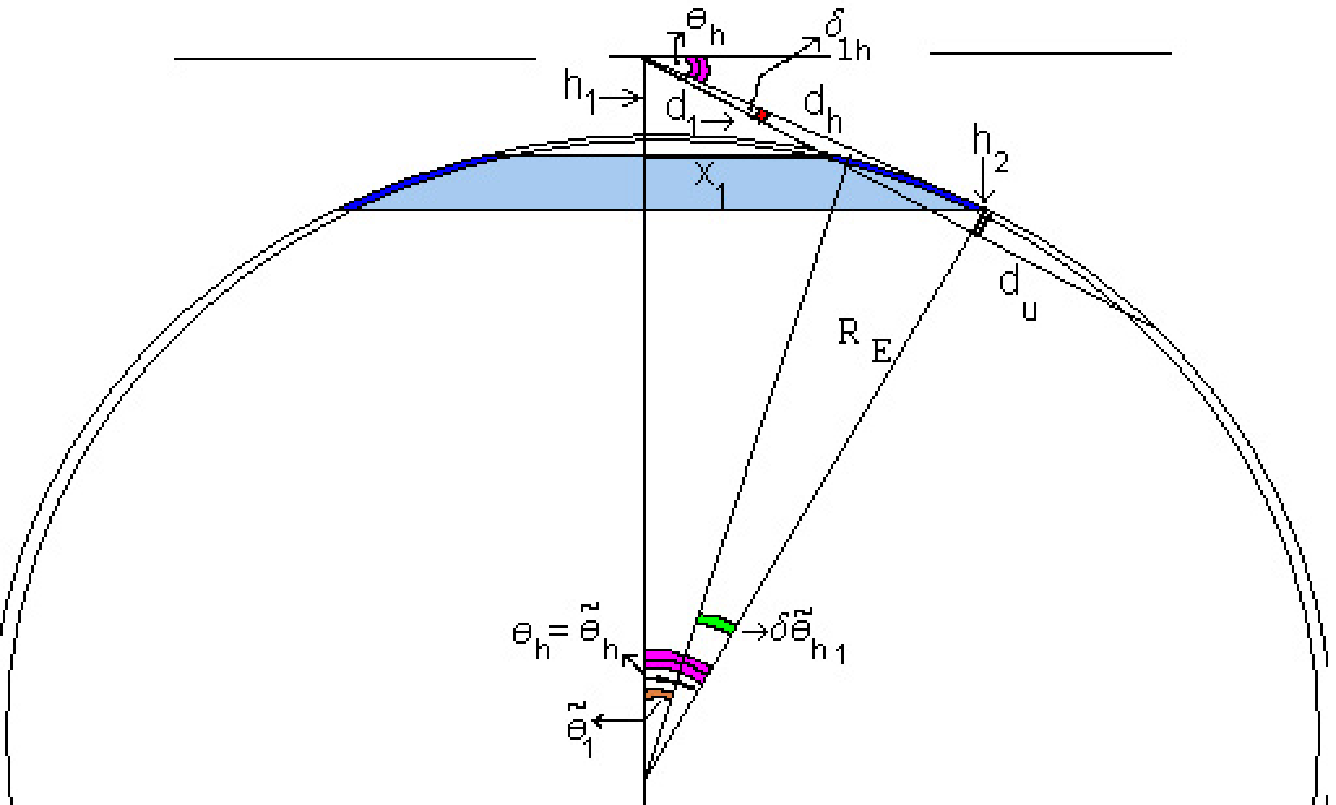}
\end{minipage}
\caption {Left:The Upward Tau Air-Shower, UPTAUs,and the
Horizontal Tau Air-Shower, HORTAU, flashing toward an observer at
height $h_1$. The HORTAU areas are described for water and rock
matter density.} \label{fig:fig1}
 \caption {Center: As previous figure, The Upward Tau
Air-Shower, UPTAUs, HORTAU, flashing toward an observer at height
$h_1$ from the Skin Crown Earth Crust seen in $3D$. The HORTAU
areas are described for water and rock matter density
\cite{Fargion 2002}.} \label{fig:fig1} \caption {Right:The
lateral geometrical disposal of the main parameters in the text
defining the UPTAUs and HORTAUs Areas; the distances are
exaggerated for simplicity.} \label{fig:fig2}
\end{figure}
%  \item
  The consequent characteristic lepton $\tau$ energy
  $E_{\tau_{h}}$ making decay  $\tau$ in flight from  $d_{h}$ distance just nearby
  the source:
%\begin{displaymath}
 ${E_{\tau_{h}}}= {\left(\frac {d_{h}}{c\tau_{0}}\right)}m_{\tau} c^2
 \simeq{2.2\cdot10^{18}eV}\sqrt{\frac {h_{1}}{km}}\sqrt{1 + \frac
{h_{1}}{2R}}.$
%\end{equation}
At low quota ($h_1 \leq$ a few kms) the  air depth before the Tau
decay necessary to develop a shower correspond to a Shower
distance $d_{Sh}$ $\sim 6 kms \ll d_h$.  More precisely at low
quota ($h_1\ll h_o$: $h_0 = 8.55$ km):
%\begin{equation}
$d_{Sh}\simeq 5.96km[ 1+ \ln {\frac{E_{\tau}}{10^{18}eV}})] \cdot
e^{\frac{h_1}{h_o}}.$
%\end{equation}
  So we may neglect the distance of the final shower respect to the
longest horizons ones. However at high  altitude ($h_1\geq h_o$)
this is no longer the case . Therefore  we shall introduce from
here and in next steps a small, but important modification ,
whose physical motivation is just to include  the air dilution
role at highest quota: $ {h}_1 \rightarrow \frac {h_{1}}{1 +
h_1/H_o}$, where , as in Appendix A, $H_o= 23$ km. Therefore
previous definition becomes:
%\begin{equation}
 ${E_{\tau_{h}}}\simeq{2.2\cdot10^{18}eV}\sqrt{\frac {h_{1}}{1 + h_1/H_o}}\sqrt{1 + \frac
{h_{1}}{2R}}.$
%\end{equation}
This procedure, applied tacitly everywhere, guarantees that there
we may extend our results to those HORTAUs at altitudes where the
residual air density  must exhibit a sufficient slant depth. For
instance, highest $\gg 10^{19} eV$ HORTAUs will be not easily
observable because their ${\tau}$ life distance exceed (usually)
the horizons air depth lenghts. The parental UHE
$\nu_{\tau}$,$\bar\nu_{\tau}$ or $\bar\nu_{e}$ energies
$E_{\nu_{\tau}}$ able to produce such UHE $E_{\tau}$ in matter: $
E_{\nu_{\tau}}\simeq 1.2 {E_{\tau_{h}}}\simeq {2.64
 \cdot 10^{18}eV \cdot \sqrt{\frac {h_{1}}{km}}}$
%\end{equation}
%\item
The neutrino (underground) interaction lenghts  at the
corresponding energies is $L_{\nu_{\tau}}$: $L_{\nu_{\tau}}=
\frac{1}{\sigma_{E\nu_{\tau}}\cdot N_A\cdot\rho_r} =
2.6\cdot10^{3}km\cdot \rho_r^{-1}{\left(\frac{E_{\nu_h}}{10^8
\cdot GeV}\right)^{-0.363}}$
%\end{displaymath}
%\begin{equation}
${\simeq 304 km
\cdot\left(\frac{\rho_{rock}}{\rho_r}\right)}\cdot{\left(\frac{h_1}{km}\right)^{-0.1815}}.
$
%\end{equation}
 For more details see \cite{Gandhi et al. 1998}, \cite{Fargion 2000-2002}, \cite{Fargion 2002}.
%\item
The maximal neutrino depth $h_{2}(h_{1})$ under the chord along
the UHE neutrino-tau trajectory of lenght $L_{\nu}(h_{1})$:
$h_{2}(h_{1}) = {\frac{L_{\nu_{h}}^2}{2^2\cdot{2}(R-h_{2})}
\simeq\frac{L_{\nu_{h}}^2}{8R_{\oplus}}}\simeq
$
%\end{displaymath}
%\begin{equation}
${\simeq 1.81\cdot km
\cdot{\left(\frac{h_1}{km}\right)^{-0.363}\cdot
\left(\frac{\rho_{rock}}{\rho_r}\right)^2}}
$
%\end{equation}
See figure above for more details. Because the above $h_2$ depths
are in general not  too deep respect to the Ocean depths, we shall
consider either sea (water) or rock (ground) materials as Crown
matter density.
%\item
The corresponding opening angle observed from height $h_{1}$,
$\delta_{1h}$ encompassing the underground height  $h_{2}$ at
horizons edge (see Fig.2) and the nearest UHE $\nu$ arrival
directions $\delta_{1}$ is \cite{Fargion 2002}:
%\begin{displaymath}
${{\delta_{1h}}(h_{2})}={2\arctan{\frac{h_{2}}{2 d_{h}}}}
={2\arctan\left[\frac{{8\cdot
10^{-3}}\cdot{(\frac{h_{1}}{km}})^{-0.863}\left(\frac{\rho_{rock}}{\rho_r}\right)^2}{{\sqrt{1+{\frac{h_{1}}{2R}}}}}\right]}
$ ${\simeq 0.91^{o}
\left(\frac{\rho_{rock}}{\rho_r}\right)^2}\cdot{(\frac{h_{1}}{km}})^{-0.863}
$
%\end{equation}
%\item
The underground chord $d_{u_{1}}$ (see Fig.$8-9$) where UHE
$\nu_{\tau}$ propagate and the nearest distance $d_{1}$ for
$\tau$ flight (from the observer toward Earth) along the same
$d_{u_{1}}$ direction, within the angle $\delta_{1h}$ defined
above, angle below the horizons (within the upward UHE neutrino
and HORTAUs propagation line) is \cite{Fargion 2002}:
%\begin{equation}
$d_{u_{1}}=2\cdot{\sqrt{{\sin}^{2}(\theta_{h}+\delta_{1h})(R_{\oplus}+{h_{1}})^{2}-{d_{h}}^2}}
$. Note that by definition  and by construction: $ L_{\nu} \equiv
d_{u_{1}} $. The nearest HORTAUs distance corresponding to this
horizontal edges still transparent to UHE $\tau$
is:${d_{1}(h_{1})}=(R_{\oplus}+h_{1})\sin(\theta_{h}+\delta_{1h})-{\frac{1}{2}}d_{u_{1}}$
Note also that for height $h_{1}\geq km$
:$\frac{d_{u_{1}}}{2}\simeq{(R_{\oplus}+{h_{1}})\sqrt{\delta_{1h}\sin{2\theta_{h}}}}\simeq
{158\sqrt{\frac{\delta_{1h}}{1^o}}\sqrt{\frac{h_{1}}{km}}}km $
%%%%%%%%%%%%%%%%%%%%%%%%%%%%%%%%%%%%%%%%%%%%%%%%%%%%%%%%%%%%%%%%%%%%%%%%%%%
%\begin{figure}\centering\includegraphics[width=8cm]{Fig04.eps}
%\caption {Distances from the observer to the Earth ($d_1$) or to
%the Horizons ($d_h$)} \label{fig:fig4}
%\end{figure}
%%%%%%%%%%%%%%%%%%%%%%%%%%%%%%%%%%%%%%%%%%%%%%%%%%%%%%%%%%%%%%%%%%%%%%%%%%%%%
%\begin{figure}\centering\includegraphics[width=8cm]{Fig03.eps}
%\caption {Distances from the observer to the Earth ($d_1$) for
%different matter densities or to the Horizons ($d_h$) for low
%altitudes} \label{fig:fig3}
%\end{figure}

%%%%%%%%%%%%%%%%%%%%%%%%%%%%%%%%%%%%%%%%%%%%%%%%%%%%%%%%%%%%%%%%%%%%%%%%%%%%%%%
%\item
The same distance projected cord $x_{1}$ $(h_{1})$ along the
horizontal line (see Fig.9):
$x_{1}(h_{1})=d_{1}(h_{1})\cos({\theta_{h}+\delta_{1h}})$ The
total terrestrial underneath any observer at height $h_{1}$ is
$A_{T}$: $=2\pi{R_{\oplus}}^{2}(1-\cos{\tilde{\theta}_{h}})
=2\pi{R_{\oplus}}h_{1}\left({\frac{1}{1+\frac{h_{1}}{R_{\oplus}}}}\right)
$;
$A_{T}=4\cdot{10}^{4}{km}^{2}{\left({\frac{h_{1}}{km}}\right)}{\left({\frac{1}{1+{\frac{h_{1}}{R}}}}\right)}
$.
%\end{equation}
Where $\tilde{\theta}_{h}$ is the opening angle from the Earth
along the observer and the horizontal point whose value is the
maximal observable one. At first sight one may be tempted to
consider all the Area  $A_{T}$ for UPTAUs and HORTAUs but because
of the air opacity (HORTAUs) or for its paucity (UPTAUs) this is
incorrect.  While for HORTAUs there is a more complex Area
estimated above and in the following, for UPTAUs the Area Ring (or
Disk) is quite simpler to derive following very similar
geometrical variables summirized in Appendix. The Earth ring
 crown crust area ${A_{R}}(h_{1})$ delimited by the horizons distance
$d_{h}$ and the nearest distance $d_{1}$ still transparent to UHE
$\nu_{\tau}$. The ring area ${A_{R}}(h_{1})$ is computed from the
internal angles $\delta{\tilde{\theta}_{h}}$ and
$\delta{\tilde{\theta}_{1}}$ defined at the Earth center (note
that $\delta{\tilde{\theta}_{h}}={\delta{\theta_{h}}}$ but in
general $\delta{\tilde{\theta}_{1}}\neq{\delta{\theta_{1}}}$).
${A_{R}}(h_{1})=2\pi{R_{\oplus}}^2(\cos{\tilde{\theta}_{1}}-\cos{\tilde\theta_{h}})
$
%\end{equation}
$=2\pi{R_{\oplus}^{2}}{\left({{\sqrt{1-{\left({\frac{x_{1}({h_1})}{R_{\oplus}}}\right)^{2}}}}-{\frac{R_{\oplus}}{R_{\oplus}+{h_{1}}}}}\right)}
$
%\end{equation}
 Here $x_{1}({h_1})$ is the cord defined above \cite{Fargion 2002}.
%%%%%%%%%%%%%%%%%%%%%%%%%%%%%%%%%%%%%%%%%%%%%%%%%%%%%%%%%%%%%%%%

%[height=0.95\textwidth,]

%%\begin{figure}\centering\includegraphics[width=8cm]{Fig04.eps}
%\%%caption {Areas and Angles for UPTAUS-HORTAUS } \label{fig:fig4}
%%\end{figure}
%%%%%%%%%%%%%%%%%%%%%%%%%%%%%%%%%%%%%%%%%%%%%%%%%%%%%%%%%%%%
\begin{figure}
\centering
\includegraphics[width=8cm]{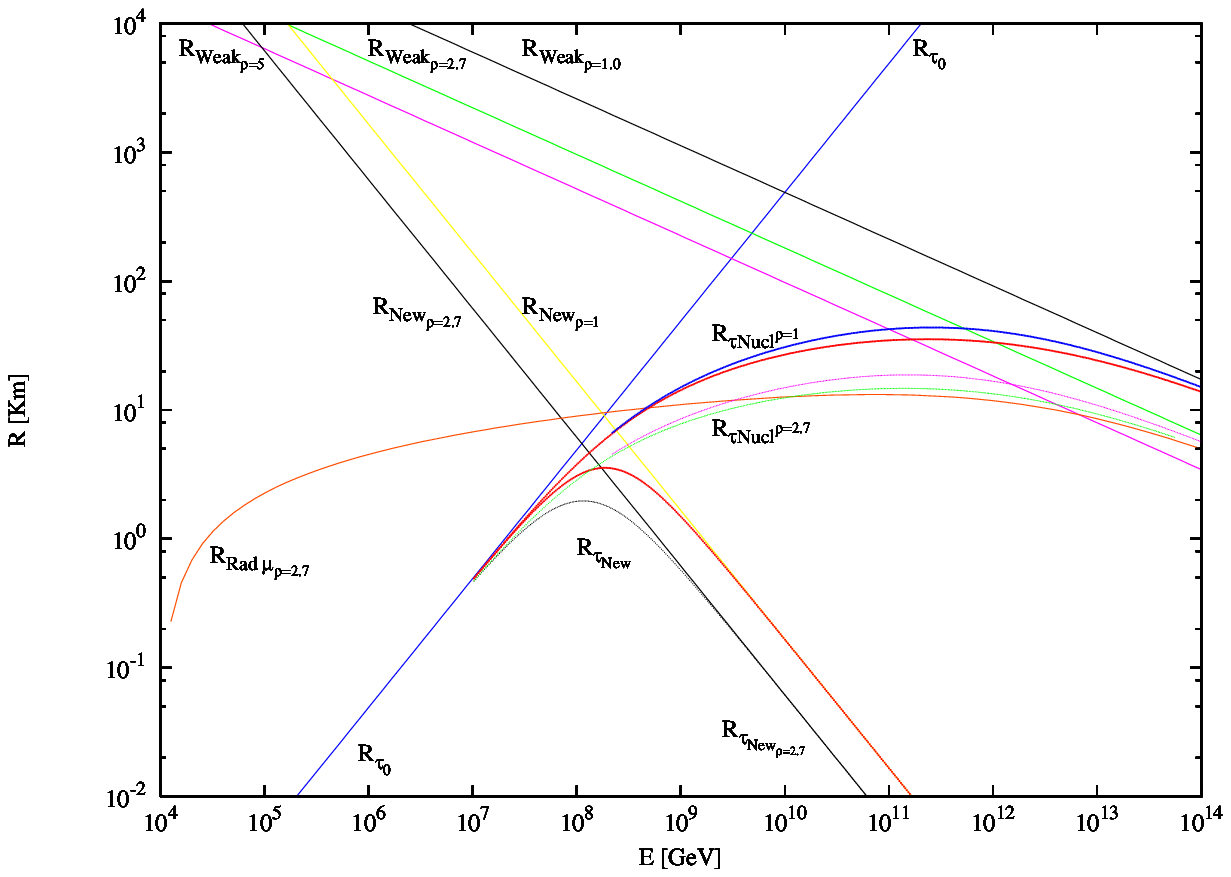}
%%%%%%%%%%%%%%%%%%%%%%%%%%%%%%%%%%%%%%%%%%%%%%%%
%\begin{figure} \centering
\includegraphics[width=8cm]{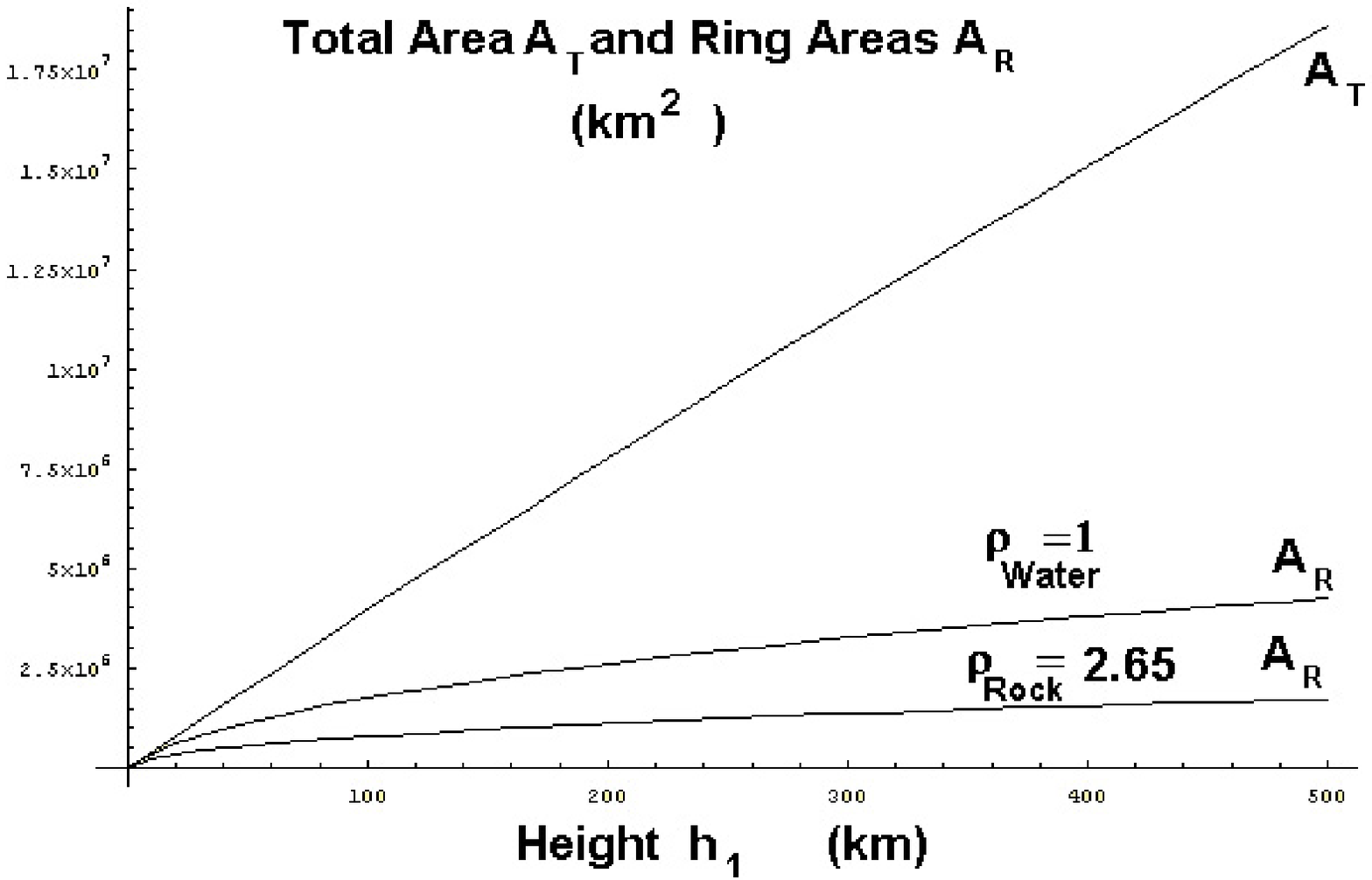}
%\end{figure}
%%%%%%%%%%%%%%%%%%%%%%%%%%%%%%%%%%%%%%%%%%%%%
\caption {Left: Lepton $\tau$ (and $\mu$) Interaction Lenghts for
different matter density: $R_{\tau_{o}}$ is the free $\tau$
lenght,$R_{\tau_{New}}$ is the New Physics TeV Gravity
interaction  range at corresponding densities,
$R_{\tau_{Nucl}\cdot{\rho}}$, \cite{Fargion 2000-2002}, see also
\cite{Becattini Bottai 2001}, \cite{Dutta et
al.2001},\cite{Fargion 2002}, is the combined $\tau$ Ranges
keeping care of all known interactions and lifetime and mainly
the photo-nuclear one. There are two slightly different split
curves (for each density) by two comparable approximations in the
interaction laws. $R_{Weak{\rho}}$ is the electro-weak Range at
corresponding densities (see also \cite{Gandhi et al.
1998});\cite{Fargion 2000-2002}\cite{Fargion 2002}.}
\label{fig:fig5}

\caption {Total Area $A_T$ and Ring Areas for two densities $A_R$
at high altitudes \cite{Fargion 2002} } \label{fig:fi8}

\end{figure}
%%%%%%%%%%%%%%%%%%%%%%%%%%%%%%%%%%%%%%%%%%%%%%%%%%%%%%%%%%%%%%%%%%%%%%%%%%%%%%%%%%
%\item
The characteristic interaction lepton tau lenght $l_{\tau}$
defined at the average $E_{\tau_{1}}$, from interaction in matter
(rock or water). These lenghts have been derived by a analytical
equations keeping care of the Tau lifetime, the photo-nuclear
losses, the electro-weak losses \cite{Fargion
2000-2002}\cite{Fargion 2002}. See figure 10 below.

%\item
The $l_{\tau_{\downarrow}}$ projected along the
$\sin(\delta\tilde{\theta}_{h_{1}})$ is defined by:
%\begin{equation}
$\delta\tilde{\theta}_{h_{1}}\equiv \tilde\theta_{h}
-\arcsin{\left({\frac{x_{1}}{R_{\oplus}}}\right)} $
%\end{equation}
%%%%%%%%%%%%%%%%%%%%%%%%%%%%%%%%%%%%%%%%%%%%%%%%%%%%%%%%%%
%\begin{figure}\centering\includegraphics[width=8cm]{Fig06.eps}
%\caption {The $\delta\tilde{\theta}_{h_{1}}$ opening angle toward
%Ring Earth Skin for density $\rho_{water}$  and $\rho_{rock}$ }
%\label{fig:fig6}
%\end{figure}

%%%%%%%%%%%%%%%%%%%%%%%%%%%%%%%%%%%%%%%%%%%%%%%%%%%%%
The same quantity in a more direct approximation:
 $\sin\delta{\tilde{\theta}_{h_{1}}}\simeq\frac{L_{\nu}}{2R_{\oplus}}=\frac{{304}km}{2R_{\oplus}}{\left({\frac{\rho_{rock}}{\rho}}\right)}{\frac{h_{1}}{km}}^{-0.1815}.
$
%\end{displaymath}
From highest ($h\gg H_o$=23km) altitude the exact approximation
reduces to:
$\delta{\tilde{\theta}_{h_{1}}}\simeq{1}^o{\left({\frac{\rho_{rock}}{\rho}}\right)}{\frac{h_{1}}{500\cdot
km}}^{-0.1815}.$
%\end{equation}
Therefore the penetrating $\tau$ skin depth
$l_{\tau_{\downarrow}}$ is: $
l_{\tau_{\downarrow}}=l_{\tau}\cdot\sin\delta{\tilde{\theta}_{h_{1}}}$,
%\begin{equation}
$\simeq{{0.0462\cdot
l_{\tau}{\left({\frac{\rho_{water}}{\rho}}\right)}}}{\frac{h_{1}}{km}}^{-0.1815}
$. Where the $\tau$ ranges in matter, $l_{\tau}$ has been
calculated and shown in Fig.10.
%%%%%%%%%%%%%%%%%%%%%%%%%%%%%%%%%%%%%%%%%%%%%%%%
%\begin{figure}\centering\includegraphics[width=8cm]{Fig07.eps}
%\caption {Total Area $A_T$ and Ring Areas for two densities $A_R$
%at low altitudes }\label{fig:fig7}
%\end{figure}
%%%%%%%%%%%%%%%%%%%%%%%%%%%%%%%%%%%%%%%%%%%%%
The final  analytical expression for the Earth Crust Skin Volumes
and Masses under the Earth Skin inspected by HORTAUs are derived
combining the above functions on HORTAUs Areas  with the previous
lepton Tau $l_{\tau_{\downarrow}}$ vertical depth depths:
%\begin{equation}
${V_{h_{1}}}={A_{R}}(h_{1})\cdot l_{\tau_{\downarrow}};
$
%\end{equation}
%\begin{equation}
${M_{h_{1}}}={V_{h_{1}}}\cdot{\left({\frac{\rho}{\rho_{water}}}\right)}
$.
%\end{equation}
%\item
%A More approximated but easy to handle
% expression for Ring area for high altitudes ($h_1\gg 2km$ $h_1\ll R_{\oplus}$) may be
%summirized as:
%\begin{displaymath}
%{A_R (h_1)}\simeq
%2\pi{R_{\oplus}^2}\sin{\theta_{h}}{\delta{\tilde{\theta}_{{h}_{1}}}}\propto{\rho^{-1}}
%\end{displaymath}
%\begin{equation}
%\simeq{2\pi{R_{\oplus}^{2}}{\sqrt{\frac{2h_{1}}{R_{\oplus}}}}\left({\frac{\sqrt{1+{\frac{h_{1}}{2R_{\oplus}}}}}{1+{\frac{h_{1}}{R}}}}\right)}{\left({\frac{L_{\nu}}{2R_{\oplus}}}\right)}
%\end{equation}
 At high altitudes the above approximation corrected accordingly to the exact one \cite{Fargion 2002}, shown in Figure
11 , becomes:
%\begin{displaymath}
${A_R (h_1)}\simeq
2\pi{R_{\oplus}}{d_{h1}}{\delta{\tilde{\theta}_{{h}_{1}}}}\simeq
{4.65 \cdot 10^6{\sqrt {\frac{h_{1}}{500
km}}}}{\left({\frac{\rho_{water}}{\rho}}\right)}{km}^2$.
%\end{equation}
% Within the above
%approximation the final searched Volume ${V_{h_{1}}}$ and Mass
 The ${M_{h_{1}}}$ from where HORTAUs may be generated is derived as:
${V_{h_{1}}}={\frac{\pi}{2}{\sqrt{\frac{2h_{1}}{R_{\oplus}}}}
\left({\frac{\sqrt{1+{\frac{h_{1}}{2R}}}}{1+{\frac{h_{1}}{R}}}}\right)}
{L_{\nu}^{2}}{l_{\tau}}\propto{\rho^{-3}}.$
%\end{equation}
%%%%%%%%%%%%%%%%%%%%%%%%%%%%%%%%%%%%%%%%%%%%%%%%%%%%%%%%%%%%%%%%%%%%%%%%%%%%%%%%%%%%%%%%%%%
%\begin{equation}
Therefore
${M_{h_{1}}}={\frac{\pi}{2}{\sqrt{\frac{2h_{1}}{R_{\oplus}}}}\left({\frac{\sqrt{1+{\frac{h_{1}}{2R}}}}{1+{\frac{h_{1}}{R}}}}\right)}{L_{\nu}^{2}}{l_{\tau}}{\rho}\propto{\rho^{-2}}.
$ The effective observable Skin Tau Mass $\Delta M_{eff.}(h_{1})$
within the thin HORTAU or UPTAUs Shower angle beam $\simeq$ $1^o$
is suppressed by the solid angle of view:
${\frac{\delta\Omega}{\Omega}} \simeq 2.5\cdot 10^{-5}$. Therefore
${\Delta
M_{eff.}(h_{1})={V_{h_{1}}}\cdot{\left({\frac{\rho}{\rho_{water}}}\right){\frac{\delta\Omega}{\Omega}}}}
$.
%\end{equation}
The Masses $\Delta M_{eff.}(h_{1})$ for realistic high quota
experiment are summirized in Table below \cite{Fargion 2002},
their consequent event rate are discussed in next section and in
the Conclusion below.
%%%%%%%%%%%%%%%%%%%%%%%%%%%%%%%%%%%%%%%%Table%%%%%%%%%%%%%%%%%%%%%%%%%%%%%%%%%%%%%%%%%%%%%%%%%%%%%%
\begin{figure}[h] \centering
\includegraphics[height=0.95\textwidth,angle=270]{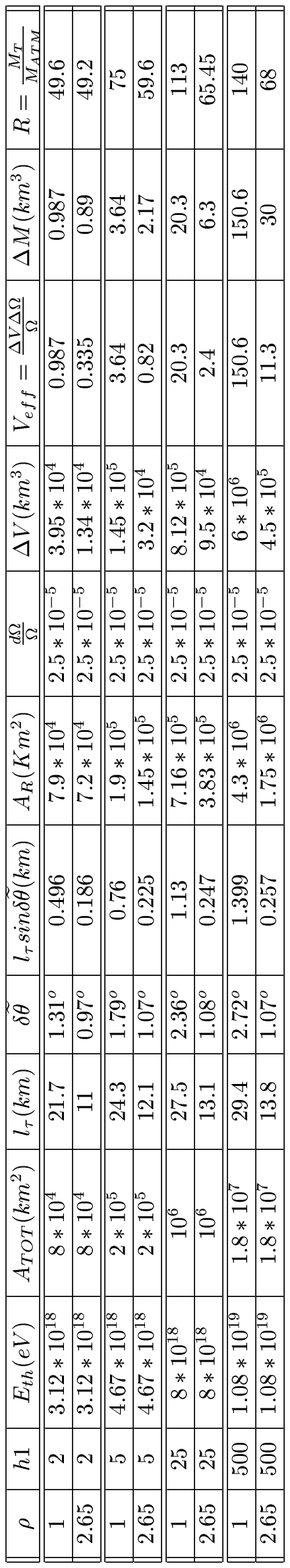}
%\includegraphics[width=7.5 cm,angle=270]{tabfargionnew2.ps}
% \caption{Horizontal Tau Air-Shower HORTAUs at different height
% $h_1$ of observation and the corresponding Energies $E_{\tau}$,
% Areas, Volumes underneath} \label{fig:fi9}
\caption{The Table of the main parameters leading to the
effective HORTAUs Mass  from the observer height $h_1$, the
corresponding $\tau$ energy $E_{\tau}$ able to let the $\tau$
reach him from the horizons, the Total Area $A_{TOT}$ underneath
the observer, the corresponding $\tau$ propagation lenght in
matter $l_{\tau}$, the opening angle toward the crown from the
Earth $\delta\tilde{\theta}_{h_1}$
 and  $l_{\tau}$ just orthogonal in the matter $l_{\tau_{\downarrow}}=l_{\tau}\cdot\sin\delta{\tilde{\theta}_{h_{1}}}$, the Ring Areas for
two densities $A_R$ at characteristic high altitudes $h_1$, the
corresponding effective Volume $V_{eff.}$ and the consequent Mass
$\Delta M_{eff.}$ (within the narrow $\tau$ Air-Shower solid
angle) as a function of density $\rho$ and height$h_1$. In the
last Column the Ratio R $= M_T/M_{ATM}$ define the ratio of
HORTAUs produced within the Earth Crown Skin over the atmospheric
ones: this ratio nearly reflects the matter over air density and
it reaches nearly two order of magnitude \cite{Fargion 2002}.}
\end{figure}
%%%%%%%%%%%%%%%%%%%%%%%%%%%%%%%%%%%%%%%%%%%%%End Table%%%%%%%%%%%%%%%%%%%%%%%%%%%%%%%%%%%%%%%%%%
\section{Event Rate for UPTAUS and HORTAUS}
The event rate for HORTAUs are given by the following expression
normalized to any given neutrino flux ${\Phi_{\nu}}$:
\begin{equation}
{\dot{N}_{year}}={\Delta
M_{eff.}}\cdot{\Phi_{\nu}}\cdot{\dot{N_o}}\cdot\frac{\sigma_{E_{\nu
}}}{\sigma_{E_{\nu_o}}}
\end{equation}
Where the ${\dot{N_o}}$ is the UHE neutrino rate estimated for
$km^3$ at any given (unitary) energy ${E_{\nu_o} }$, in absence of
any Earth shadow. In our case we shall normalize our estimate at
${E_{\nu_o}=3}$ PeVs energy for standard electro-weak charged
current in a standard parton model \cite{Gandhi et al. 1998} and
we shall assume a  model-independent neutrino maximal flux
${\Phi_{\nu}}$ at a flat fluence value of nearly ${\Phi_{\nu}}_o$
$\simeq 3\cdot 10^3 eV cm^{-2}\cdot s^{-1}\cdot sec^{-1}\cdot
sr^{-1}$ corresponding to a characteristic Fermi power law in UHE
$\nu$ primary production rate decreasing as $\frac
{dN_{\nu}}{dE_{\nu}}\simeq {E_{\nu}}^{-2}$ just below present
AMANDA bounds. The consequent rate becomes: $ {\dot{N}_{year}}=
29 {\frac{{\Delta
M_{eff.}}}{km^{3}}\cdot\frac{{\Phi_{\nu}}}{{\Phi_{\nu}}_o}}\cdot\frac{\sigma_{E_{\nu
}}}{\sigma_{E_{\nu_o}}} $
%\end{displaymath}
\begin{equation}
{\dot{N}_{year}}= 29\cdot{\left(\frac{E_{\nu}}{3 \cdot 10^6 \cdot
GeV}\right)^{-0.637}} {\frac{{\Delta
M_{eff.}}}{km^{3}}\cdot\frac{{\Phi_{\nu}}}{{\Phi_{\nu}}_o}}
\end{equation}
For highest satellites and for a characteristic UHE GZK energy
fluence ${\Phi_{\nu}}_o$  $\simeq 3 10^3 eV cm^{-2}\cdot
s^{-1}\cdot sr^{-1}$ (as the needed Z-Showering one), the
consequent event rate observable ${\dot{N}_{year}}$ above the Sea
is from satellite quota ({500 km}):
\begin{equation}
{\dot{N}_{year}}=12.3\cdot{\left(\frac{E_{\nu}}{3 \cdot 10^{10}
\cdot GeV}\right)^{-0.637}}
\cdot\frac{{\Phi_{\nu}}}{{\Phi_{\nu}}_o}
\end{equation}
 This event rate is comparable to UPTAUS one and it may be an
 additional source of Terrestrial Gamma Flashes observed by GRO
 in last decade \cite{Fargion 2000-2002},\cite{Fargion 2002}.
%%%%%%%%%%%%%%%%%%%%%%%%%%%%%%%%%%%%%%%%%%%%%%%%%%%%%%%%%%%%%%%%%%%%%%%%%%%%%%%%%%%5
%%%%%%%%%%%%%%%%%%%%%%%%%%%%%%%%%%%%%%%%%%%%%%%%%%%%%%%%%%%%%%%%%%%%%%%%%%%%%%%%%%%%%%%%%
\section{Summary and Conclusions}
The discover of the expected UHE neutrino Astronomy is urgent and
just behind the corner. Huge volumes are necessary. Beyond
underground $km^3$ detectors a new generation of UHE neutrino
calorimeter lay on front of mountain chains and just underneath
our feet: The Earth itself  offers huge Crown Volumes as Beam
Dump calorimeters observable via upward Tau Air Showers, UPTAUs
and HORTAUs. Their effective Volumes as a function of the quota
$h_1$ has been derived by an analytical function variables in
equations above  and Appendix. These Volumes and Masses are
discussed below and summirized in the last column of the above
Table; they are  large enough to offer  an ideal calorimeter for
future UHE neutrino detection.

%%%%%%%%%%%%%%%%%%%%%%%%%%%%%%%%%%%%%%%%%%%%%%%%%%%%%%%%%%%%%%%%%%%%%%%%%%%%%%
%%%%%%%%%%%%%%%%%%%%%%%%%%%%%%%%%%%%%%%%%%%%%%%%%%%%%%%%%%%%%%%%%%%%%%%%%%%%%%%%%%%

%%%%%%%%%%%%%%%%%%%%%%%%%%%%%%%%%%%%%%%%%%%%%%%%%%%%%%%%%%%%%%%%%%%%%%%%%%%%%%%%

\section{Appendix : The UPTAUS Area}
The Upward Tau Air-Showers, mostly at PeV energies, might travel a
minimal air depth before reaching the observer in order to
amplify its signal. The UPTAUS Disk Area $A_U$ underneath an
observer at height $h_1$ within a opening angle $\tilde{\theta}_2$
from the Earth Center is:
%\begin{equation}
$ A_{U}= 2\pi{R_{\oplus}}^2(1 - \cos{\tilde{\theta}_{2}}).
%\end{equation}
$
Where the $\sin{\tilde{\theta}_{2}= (x_2/{R_{\oplus}})}$ and
$x_2$ behaves like $x_1$ defined above for HORTAUs. In general the
UPTAUs area are constrained in a narrow Ring (because the mountain
presence itself or because the too near observer distances from
Earth are encountering a too short air slant depth for showering
or a too far and opaque atmosphere for the horizontal UPTAUs):
%\begin{equation}
 $A_{U}= 2\pi{R_{\oplus}}^2( \cos{\tilde{\theta}_{3}-\cos{\tilde{\theta}_{2}}})
$
 %\end{equation}
An useful Euclidean approximation is: $ A_{U}= \pi {h_1}^2
({\cot{\theta}_{2}}^2-{\cot{\theta}_{3}}^2)
 $
 Where ${\theta}_{2}$, ${\theta}_{3}$ are the outgoing $\tau$
 angles on the Earth surface \cite{Fargion 2000-2002},\cite{Fargion 2002}.

\end{document}